\documentclass[twocolappendix, twocolumn]{aastex631}

\usepackage{amsmath}
\usepackage{makecell}
\usepackage{appendix}
\usepackage{multirow}
\usepackage{colortbl}

\newcommand{\nc}[1]{\textcolor[rgb]{0.0, 0.0, 0.0}{#1}}

\begin{document}

\title{Precision Determination of Scintillation Screen Parameters from Annual Modulation Measurement of Pulsar Scintillation Arc Curvature with the FAST Telescope}

\author[0009-0006-4956-8803]{Yuanshang Huang}
\altaffiliation{huangyuanshang42@gmail.com}
\affiliation{South-Western Institute for Astronomy Research (SWIFAR), Yunnan University, 650500 Kunming, China}

\author[0000-0003-2076-4510]{Xun Shi}
\altaffiliation{xun@ynu.edu.cn}
\affiliation{South-Western Institute for Astronomy Research (SWIFAR), Yunnan University, 650500 Kunming, China}

\author[0000-0002-4997-045X]{Jumei Yao}
\affiliation{Xinjiang Astronomical Observatory, Chinese Academy of Sciences, 150 Science 1-Street, Urumqi, Xinjiang 830011, China}
\affiliation{Key Laboratory of radio Astronomy, Chinese Academy of Sciences, Urumqi, Xinjiang, 830011, China}
\affiliation{Xinjiang Key Laboratory of radio Astrophysics, 150 Science1-Street, Urumqi, Xinjiang, 830011, China}

\author[0000-0001-5105-4058]{Weiwei Zhu}
\affiliation{National Astronomical Observatories, Chinese Academy of Sciences, Beijing 100101, China}
\affiliation{Institute for Frontiers in Astronomy and Astrophysics, Beijing Normal University, Beijing 102206, China}

\author[0000-0001-5662-6254]{Yonghua Xu}
\affiliation{Yunnan Observatories, Chinese Academy of Sciences, 650216 Kunming, China}
\affiliation{Key Laboratory of Radio Astronomy and Technology, Chinese Academy of Sciences, Beijing 100101, China}




\begin{abstract}

Pulsar scintillation observations have revealed ubiquitous discrete scintillation screens in the interstellar medium. A major obstacle in identifying the nature of these screens is the uncertainty in their distances, which prevents precise correlation with known structures in the Milky Way. We used the Five-hundred-meter Aperture Spherical radio Telescope (FAST) to observe PSR B1237+25, PSR 1842+14, and PSR 2021+51. We detected 10 scintillation arcs in PSR B1237+25, 1 in PSR 1842+14, and at least 6 in PSR 2021+51. By modeling the annual modulation of these scintillation arcs, we constrained the distances of the scintillation screens, as well as the anisotropic scattering directions and the projected velocities in those directions. The scintillation screens are distributed throughout the entire paths between Earth and the pulsars. Among these, the distance to the main scintillation screen toward PSR B1237+25 is $267^{+32}_{-28}$ pc, the scintillation screen toward PSR B1842+14 is at a distance of $240^{+210}_{-120}$ pc, and the main scintillation screen toward PSR B2021+51 is located at $887^{+167}_{-132}$ pc. Several screens in our sample appear at distances coinciding with the Local Bubble boundary, particularly the brightest scintillation arc toward PSR B1237+25. We provide a substantial sample of scintillation screen measurements, revealing the rich plasma density fluctuation structures present in the Milky Way.

\end{abstract}



\section{Introduction} \label{sec:intro}

Pulsars are highly magnetized, rotating neutron stars that emit beams of radio waves from their magnetic poles \citep{1968Natur.217..709H, 1968Natur.218..731G}. Inhomogeneities in the electron density of the ionized interstellar medium (IISM) scatter a pulsar’s radio waves as they propagate, leading to multi-path interference. As the Earth, the pulsar, and the IISM move relative to each other, the interference conditions change with time. Consequently, the pulsar’s radio signal intensity is modulated in time and frequency, a phenomenon known as interstellar scintillation (ISS) \citep{1968Natur.218..920S}. Observations and analysis of pulsar scintillation provide a powerful probe of the IISM. 
\nc{Sub-structure from scintillation arcs and arclets implies spatial scales even smaller than 1 AU by a few orders of magnitude} \citep{1990ARA&A..28..561R}, making ISS an effective tool for investigating extremely small-scale structures in the IISM.

Subsequent observations have shown that interstellar scintillation is concentrated in discrete regions along the line of sight referred to as ``scintillation screens'', and that the scintillation screens are prevalent in the IISM. A survey conducted using the Green Bank Telescope (GBT) and the Arecibo Observatory \citep{Stinebring_2022} detected definite or probable scintillation screens in 19 out of 22  pulsars with low to moderate dispersion measure. \cite{Wu_2022} conducted the first sizeable census of diffractive pulsar scintillation at low frequencies (120 - 180 MHz) using the Low-Frequency Array (LOFAR), detecting scintillation screens in 9 out of the 15 pulsars that allow for a full determination of the scintillation properties. \cite{Main_2023_107P} observed 107 pulsars with interstellar scintillation screens using the MeerKAT Thousand Pulsar Array, concluding that these screens appear to be ubiquitous in clean, high signal-to-noise observations. High-sensitivity measurements have revealed multiple distinct arcs along the line of sight to individual pulsars. For instance, \cite{2024MNRAS.527.7568O} observed at least 9 separate screens toward B1929+10 with the Five-hundred-meter Aperture Spherical radio Telescope (FAST), \cite{reardon2025} used the MeerKAT radio telescope to discover 25 screens in the direction of the millisecond pulsar PSR J0437-4715.

Scintillation screens can be associated with various interstellar structures, including H II regions \citep{Mall_2022}, bow shocks \citep{reardon2025}, and supernova remnants \citep{2021NatAs...5..788Y}. Establishing connections between scintillation screens and these structures can reveal the origin of the scintillation screens while simultaneously providing insights into both the structures themselves and the physical properties of the associated pulsars. For instance, \cite{2021NatAs...5..788Y} discovered a scintillation screen associated with the supernova remnant, and by utilizing this screen as a probe, they obtained the first evidence that the three-dimensional direction of motion in a young pulsar aligns with its rotation axis. \nc{Additionally, scattering associated with the Crab pulsar has been extensively studied since the pulsar's discovery \citep{Staelin1968}, with detailed characterization of nebular scattering (e.g., \citealt{Cordes_2004}, \citealt{McKee2018}, \citealt{Main_2021}, \citealt{Lin_2023}, \citealt{Serafin_2024}).} Another example is that \cite{reardon2025} linked four scintillation screens to the bow shock, and through analysis of these screens, they revealed details about the bow shock structure as well as the pulsar's radial velocity.

However, in most cases, the astrophysical associations of scintillation screens are still unknown. Also under debate is the physical nature of the microscopic electron density fluctuations that generate the scintillation. There exists a promising proposal that scintillation screens are related to reconnection sheets - dissipative regions of magnetohydrodynamic (MHD) turbulence in the IISM \citep{Goldreich_Sridhar_2006, Pen_King_2012, Pen_Levin_2014, Liu_2016, Simard_Pen_2018}. A major impediment to understanding the nature of scintillation screens and discovering their potential macroscopic associations is the uncertainty in their distances. So far, most observed scintillation screens still do not have well-determined distances.

A thin-screen model is often used to describe the IISM density fluctuations responsible for the pulsar scintillation \citep{1968Natur.218..920S}. \cite{Stinebring_2001} explained that scintillation screens can produce parabolic features in the two-dimensional power spectrum, which are referred to as the scintillation arcs. The scintillation arcs can be described as a function of $f_{\nu}$ (differential time delay) and $f_{t}$ (differential Doppler shift): $f_{\nu}$ = $\eta_{\nu} f_{t}^{2}$, where $\eta_{\nu}$ is arc curvature. \cite{Walker_2004} and \cite{Cordes_2006} later presented models explaining the geometry of these scintillation arcs. Based on their now commonly accepted model, the curvature $\eta_{\nu}$ of a scintillation arc encodes information on the distance to the scintillation screen.
 
In early times, the curvature of a single scintillation arc measurement \nc{was} used to determine the distance to the screen (e.g., \citealt{Putney_2006}). Such distance determination neglects the possible transverse velocity of the scintillation screen and assumes that the scattering is isotropic in the scintillation screen. Later observations have found a high degree of anisotropy in the scattering in some scintillation screens (e.g., \citealt{2010ApJ...708..232B}, \citealt{2022MNRAS.515.6198S}, \citealt{Stinebring_2022}). Considering this high degree of anisotropy, the parabolic arc curvature depends on three parameters of the screen: the distance from the observer, the main scattering direction, and the velocity projected along the main scattering direction. \nc{VLBI observations of pulsar scintillation allow for measurement of the scattered brightness distribution with unprecedented precision, revealing fine-scale structures in the scintillation screen, and effectively lifting the parameter degeneracy by providing phase information (\citealt{2010ApJ...708..232B}; \citealt{Stock_2025}).} 

Apart from VLBI, the only feasible method to measure these screen parameters is to use the annual modulation of parabolic arc curvature. \nc{By observing the scintillation arc curvature of a pulsar at multiple epochs throughout an observing year, one can determine the main scattering direction by measuring how Earth's orbital motion projects onto the scattering axis, then determine the screen distance from the variation in arc curvature combined with the pulsar distance, and finally determine the screen velocity from the mean arc curvature combined with the pulsar distance and proper motion} (e.g., \citealt{2020ApJ...904..104R}; \citealt{McKee_2022}; \citealt{Mall_2022}; \citealt{2023MNRAS.526.1246X}; \citealt{2023SCPMA..6619512L}).

Extended scintillation arcs are extremely faint features, so high-sensitivity radio telescopes are required to measure their curvature precisely. The Five-hundred-meter Aperture Spherical Telescope (FAST), which has an illuminated aperture of up to 300 m \citep{2019SCPMA..6259502J}, offers extremely high sensitivity and is particularly well-suited for such observations. We used FAST to observe pulsars B1237+25, B1842+14, and B2021+51. We then modeled the annual modulation of the scintillation arc curvature, taking into account the Earth’s orbital motion and the pulsar’s proper motion. From this modeling, we determined the location of the scintillation screen, as well as the main scattering direction and the velocity projected along that direction.

This paper is organized as follows: Section \ref{sec:Theory of annul modulation} discusses the theory of the annual modulation of scintillation arc curvature; Section \ref{sec:Observations and data processing} describes the observational data and data processing methods; Section \ref{results} presents the derived parameters of the scintillation screens and discusses their distribution and potential sources; Section \ref{Conclusion} provides a summary of our findings and an outlook for future research.

\section{Theory of annual modulation}
\label{sec:Theory of annul modulation}

\begin{figure*}[ht]
\centering
\includegraphics[width=\textwidth]{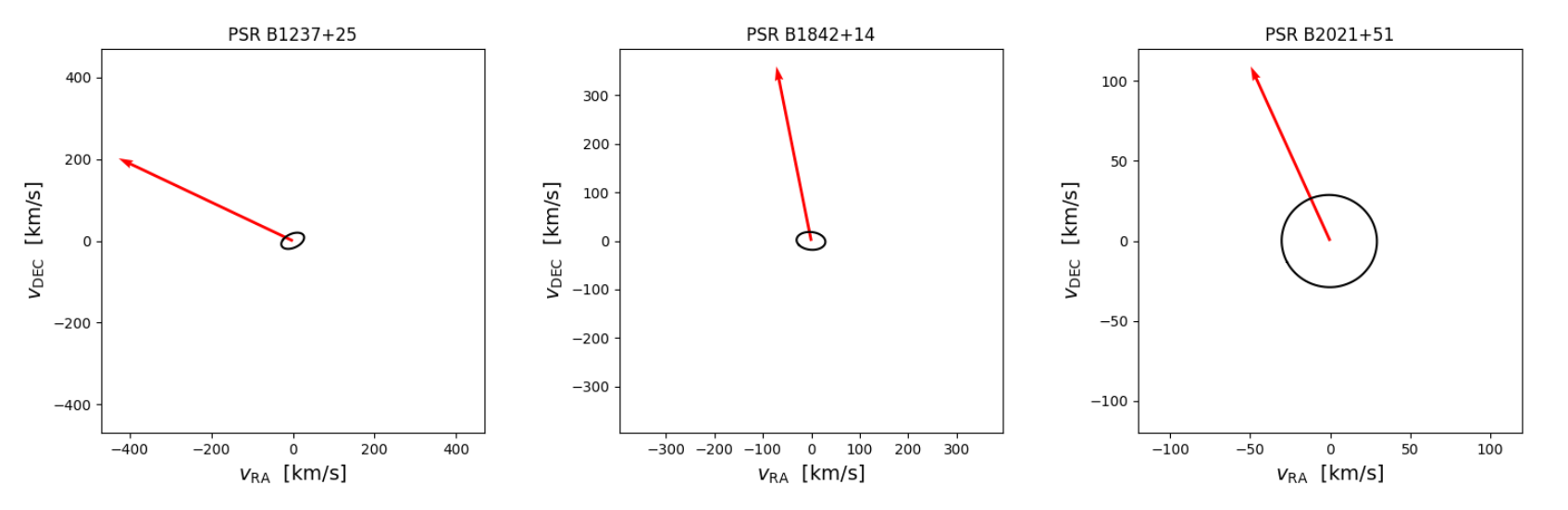}
\caption{Transverse velocity of the pulsar \( \mathbf{v}_{\text{psr}}\) (red arrow) and the projected Earth’s velocity \( \mathbf{v}_{\text{Earth}}(t)\) (black ellipse) for the three pulsars.}
\label{fig:1}
\end{figure*}

The scintillation arc curvature depends on the observing frequency \( \nu \), the pulsar distance \( d_{\text{psr}} \), the fractional distance $s$ of the scintillation screen to the pulsar, and the projected relative velocity of the screen with respect to the pulsar-observer line of sight \( v_{\text{sr}} \) (see e.g. \citealt{2010ApJ...708..232B}),

\begin{equation}
\eta_{\nu} = \frac{c d_{\text{psr}} s(1 - s)}{2 \nu^2 v_{\text{sr}}^2} \,,
\end{equation}

\noindent where $c$ is the speed of light.

When the scattering is isotropic in the screen,

\begin{equation}
v_{\text{sr}} = \left| s \mathbf{v}_{\text{Earth}} + (1 - s)\mathbf{v}_{\text{psr}} - v_{\text{screen}} \mathbf{n}_{\text{screen}} \right|.
\end{equation}

\noindent Here, all velocities are 2D transverse velocities with respect to the pulsar line of sight. \( \mathbf{v}_{\text{Earth}} \) and \( \mathbf{v}_{\text{psr}} \) are the Earth's and the pulsar's velocities, respectively. \( v_{\text{screen}} \) is the velocity of the screen and \( \mathbf{n}_{\text{screen}} \) is its 2D orientation vector in the sky.

When the scattering is highly anisotropic, the parabolic arc corresponds to the scattered images along the main scattering direction, and

\begin{equation}
v_{\text{sr}} = \left[ s \mathbf{v}_{\text{Earth}} + (1 - s)\mathbf{v}_{\text{psr}} \right] \cdot \mathbf{n}_{\text{screen}} - v_{\text{screen}}.
\label{vsr_iso}
\end{equation}

\noindent In this case, \( \mathbf{n}_{\text{screen}} \) stands for the main scattering direction and \( v_{\text{screen}} \) is the screen velocity along this direction. We shall limit \( \mathbf{n}_{\text{screen}} \) to be within [0, 180) degrees of a specified direction.

In the case of the isolated pulsars, the modulation of \(\eta_{\nu}\) results from the change of \( \mathbf{v}_{\text{Earth}} \) as the Earth orbits around the Sun. Note that \( \eta_{\nu}^{-1/2} \) is linearly related to the velocities. Taking advantage of this and separating the measurement and the model like in \cite{Mall_2022} and \cite{2022MNRAS.515.6198S}, one obtains

\begin{equation}
|f_{\eta}| = \left( \frac{\eta_{\nu} \nu^2}{c} \right) ^{-\frac{1}{2}}
\end{equation}

\noindent with

\begin{equation}
f_{\eta} \equiv \frac{\sqrt{2} v_{\text{sr}}}{\sqrt{s(1-s)d_{\text{psr}}}}.
\end{equation}

\noindent This serves as a convenient form of the model for the parameter fitting.

Given the distance and the proper motion of the pulsar, we are left with three unknown parameters $s$, \( v_{\text{screen}} \), and \( \mathbf{n}_{\text{screen}} \). They are to be fitted from the annual modulation of the measurements \( | f_\eta | (t) \).

\subsection{Analytical understanding of the parameter dependence}
\label{subsec:Analytical understanding of the parameter dependence}

\subsubsection{1D scattering}
\label{subsubsec:1D scattering}

Let us first consider the simpler case of 1D scattering, i.e., extremely anisotropic scattering. Approximating the Earth’s orbit as circular, \( f_{\eta} (t) \) has a sinusoidal shape that is characterized by three quantities: the modulation amplitude \( A_{\text{f}} \equiv (f_{\eta\text{, max}} - f_{\eta\text{, min}})/2 \), the mean offset \(\bar{f_\eta}\), and the yearly time \( t_{\text{max}} \) when \( f_\eta (t) \) reaches its maximum. They are dependent on the three parameters as follows.

When projected onto the direction of the pulsar, the Earth’s velocity vector \(\mathbf{v}_{\text{Earth}}(t)\) forms an ellipse around the year (see Figure \ref{fig:1}). This velocity ellipse is further projected onto the direction of \(\mathbf{n}_{\text{screen}}\) to obtain \(v_{\text{proj}} \equiv \mathbf{v}_{\text{Earth}} \cdot \mathbf{n}_{\text{screen}} \). The modulation amplitude \( A_{\text{f}} \) is set by the maximum of \(v_{\text{proj}}\),

\begin{equation}
A_{\text{f}} = \sqrt{\frac{2 s}{(1 - s)}} \frac{v_{\text{proj,max}}}{\sqrt{d_\text{psr}}},
\end{equation}

\noindent and the maximum of \( f_\eta (t) \) is reached at time \( t_\text{max} \) when \(|v_{\text{proj}}| = v_{\text{proj,max}} \). Therefore, from \( t_\text{max} \) one can determine the orientation \(\mathbf{n}_{\text{screen}}\). Given \(\mathbf{n}_{\text{screen}}\), one can compute \(v_{\text{proj,max}}\) and obtain s from the measurement of the modulation amplitude \(A_{\text{f}}\).

The mean offset \(\bar{f_\eta}\) is given by the pulsar and screen velocities,

\begin{equation}
\bar{f_\eta} = \left[ 1 - \frac{1}{1-s} \frac{v_\text{screen}}{\mathbf{v}_\text{psr} \cdot \mathbf{n}_{\text{screen}}} \right] \sqrt{\frac{2(1-s)}{s}} \frac{\mathbf{v}_\text{psr} \cdot \mathbf{n}_{\text{screen}}}{{\sqrt{d_\text{psr}}}}.
\end{equation}

\noindent It depends on all three parameters. When the other two parameters are determined from \(t_\text{max}\) and \(A_\text{f}\), one can further determine \(v_\text{screen}\) from \(\bar{f_\eta}\).

Since a physical screen velocity is typically much smaller than the pulsar proper motion velocity, one generally has \( v_{\text{screen}} / (\mathbf{v}_\text{psr} \cdot \mathbf{n}_{\text{screen}}) / (1 - s) \ll 1 \), i.e.

\begin{equation}
\bar{f_n} \approx \sqrt{\frac{2(1-s)}{s}} \frac{\mathbf{v}_\text{psr} \cdot \mathbf{n}_{\text{screen}}}{{\sqrt{d_\text{psr}}}}.
\end{equation}

\noindent This then forms a consistency test on the measurements: there must
exist a projection direction \( \mathbf{n}_{\text{screen}} \) that allows the modulation measurements to satisfy the following constraint

\begin{equation}
A_\text{f} \bar{f_\text{n}} \approx \frac{2 v_{\text{proj,max}} \mathbf{v}_\text{psr} \cdot \mathbf{n}_{\text{screen}}}{d_\text{psr}} \,,
\end{equation}
independent of the unknown fractional screen distance.

\subsubsection{2D isotropic scattering}
\label{subsubsec:2D isotropic scattering}

The full geometry of a general 2D scintillation screen cannot be constrained by the curvature measurements. It requires full modeling of the distribution of scattered power along and within a parabolic arc \citep{2020ApJ...904..104R}, and is subject to degeneracies. The ideal case of 2D isotropic scattering, however, can be fully captured with a three-parameter model similar to the case of 1D scattering. In the early scintillation studies, scattering was thought to be produced by statistically isotropic ISM turbulence, and thus a 2D isotropic scattering was widely used. Deviation of the shape of \( f_\eta (t) \) from a sinusoidal shape can be a sign of 2D scattering. For 2D isotropic scattering, the shape of \( f_\eta (t) \) is in general not symmetric around the mean, but this asymmetric deviation from the sinusoidal shape is only strong when the velocity modulation by Earth is comparable to the yearly average of \( v_{\text{sr}} / s\).

Apart from this strong-modulation case, the analytical framework for 1D scattering still holds approximately. The only difference is the projection of \(\mathbf{v}_{\text{Earth}}\) to \(v_{\text{proj}}\) that should be performed along the direction of \( (1 - s)\mathbf{v}_\text{psr} - v_{\text{screen}} \mathbf{n}_{\text{screen}} \) instead of \( \mathbf{n}_{\text{screen}} \).

\section{Observation and Data Processing} \label{sec:Observations and data processing}
\subsection{Observation}

Our data were obtained from observations of isolated pulsars B1237+25, B1842+14, and B2021+51 between November 2019 and July 2023 using the central beam of the FAST telescope’s L-band 19-beam receiver \citep{2020RAA....20...64J}. The properties of the observed pulsars are shown in Table~\ref{tab:PSRs}. 

B1237+25 was observed 10 times, B2021+51 was observed 5 times, and B1842+14 was observed 4 times. Each observation lasted between 30 minutes and 120 minutes. The central beam of the 19-beam receiver recorded filterbank data in four polarization channels. These data have either 8192 or 4096 frequency channels covering the frequency range from 1000 MHz to 1500 MHz, corresponding to a frequency resolution of 61 kHz or 122 kHz. For polarization-related analyses, only data within the range of 1050 MHz to 1450 MHz can be used, while for analyses involving only the total intensity, the data from both 50 MHz edges can be retained. The observational setting for pulsars is shown in Table~\ref{tab:pulsar_observations}.

\renewcommand{\arraystretch}{1.5}
\setlength{\tabcolsep}{3pt}
\begin{deluxetable*}{cccccccccc}
\tablecaption{The observed pulsars and their properties. $P$ (spin period) and $\dot{P}$ (spin-down rate) are retrieved from the ATNF catalogue \citep{Manchester_2005}. All other properties of PSR B1237+25 and PSR B2021+51 were obtained from measurements using the NRAO Very Long Baseline Array \citep{2002ApJ...571..906B}. The proper motion and the location of PSR B1842+14 were obtained from a program of observations carried out at Jodrell Bank to measure the proper motions of pulsars \citep{1993MNRAS.261..113H}. The DM of PSR B1842+14 was obtained from The Thousand-Pulsar-Array program on MeerKAT \citep{2024MNRAS.530.1581K}. \cite{2020RAA....20...76Y}, using the YMW16 model \citep{2017ApJ...835...29Y}, adopted an uncertainty of 40\% of the distance (1.7 kpc) as a rough estimate for the one-sigma error of the DM-distance to PSR B1842+14. Therefore, we also adopt a distance of 1.7 ± 0.7 kpc for PSR B1842+14. The $\text{S}_{1400}$ of PSR B1842+14 was also obtained from the research by \cite{2020RAA....20...76Y}.}
\label{tab:PSRs}
\tablehead{
\colhead{Pulsar} & 
\colhead{\thead{RAJ \\ (hms)}} & 
\colhead{\thead{DECJ \\ (dms)}} & 
\colhead{\thead{Parallax (mas) or \\ Distance (kpc)}} & 
\colhead{\thead{PMRA \\ (mas yr$^{-1}$)}} & 
\colhead{\thead{PMDEC \\ (mas yr$^{-1}$)}} & 
\colhead{\thead{$P$ \\ (ms)}} & 
\colhead{\thead{$\dot{P}$ \\ (s s$^{-1}$)}} &
\colhead{\thead{$\text{S}_{1400}$ \\ (mJy)}} & 
\colhead{\thead{DM \\ ($\text{cm}^{-3}$ pc)}}
}
\startdata
B1237+25 & 12:39:40.36 & +24:53:50.01 & $1.16 \pm 0.08$ (mas) & $-106.82 \pm 0.17$ & $49.92 \pm 0.18$ & 1382 & 9.60 $\times 10^{-16}$ & 20 & 9.28 \\
\arrayrulecolor[gray]{0.6}\cline{1-10}\arrayrulecolor{black}
B1842+14 & 18:44:54.88 & +14:54:13.89 & $1.7 \pm 0.7$ (kpc) & $-9 \pm 10$ & $45 \pm 6$ & 375 & $1.87 \times 10^{-15}$ & 1.8& 41.49\\
\arrayrulecolor[gray]{0.6}\cline{1-10}\arrayrulecolor{black}
B2021+51 & 20:22:49.87 & +51:54:50.39 & $0.50 \pm 0.07$ (mas) & $-5.23 \pm 0.17$ & $11.54 \pm 0.28$ & 529 & $3.06 \times 10^{-15}$ & 27 & 22.58\\
\enddata
\end{deluxetable*}

\renewcommand{\arraystretch}{1.5}
\setlength{\tabcolsep}{5.5pt}
\begin{deluxetable*}{c|cccccc}
\tablefontsize{\footnotesize}
\tablecaption{Observational settings for pulsars B1237+25, B1842+14, and B2021+51 using the L-band 19-beam receiver at the center of the FAST telescope.}
\label{tab:pulsar_observations}
\tablehead{
\colhead{Pulsar} & \colhead{MJD} & \colhead{UTC Date (y-m-d)} & \colhead{Duration (s)} & \colhead{Number of Channels} & \colhead{Frequency Resolution (kHz)} & \colhead{Observation Mode}
}
\startdata
\multirow{10}{*}{B1237+25} & 58808.1 & 2019-11-21 & 3600 & 4096 & 122 & Tracking \\
\arrayrulecolor[gray]{0.6}\cline{2-7}\arrayrulecolor{black}
 & 58865.9 & 2020-01-17 & 7200 & 4096 & 122 & Tracking \\
 \arrayrulecolor[gray]{0.6}\cline{2-7}\arrayrulecolor{black}
 & 59231.8 & 2021-01-17 & 1800 & 4096 & 122 & Tracking \\
 \arrayrulecolor[gray]{0.6}\cline{2-7}\arrayrulecolor{black}
 & 59297.7 & 2021-03-24 & 1800 & 4096 & 122 & Tracking \\
 \arrayrulecolor[gray]{0.6}\cline{2-7}\arrayrulecolor{black}
 & 59837.2 & 2022-09-15 & 3456 & 8192 & 61 & Swift Calibration \\
 \arrayrulecolor[gray]{0.6}\cline{2-7}\arrayrulecolor{black}
 & 59899.1 & 2022-11-16 & 3600 & 8192 & 61 & Swift Calibration \\
 \arrayrulecolor[gray]{0.6}\cline{2-7}\arrayrulecolor{black}
 & 59959.0 & 2023-01-14 & 3600 & 8192 & 61 & Swift Calibration \\
 \arrayrulecolor[gray]{0.6}\cline{2-7}\arrayrulecolor{black}
 & 60021.6 & 2023-03-18 & 3600 & 8192 & 61 & Swift Calibration \\
 \arrayrulecolor[gray]{0.6}\cline{2-7}\arrayrulecolor{black}
 & 60090.5 & 2023-05-26 & 3600 & 8192 & 61 & Swift Calibration \\
 \arrayrulecolor[gray]{0.6}\cline{2-7}\arrayrulecolor{black}
 & 60143.4 & 2023-07-18 & 3600 & 8192 & 61 & Swift Calibration \\ \hline
\multirow{4}{*}{B1842+14} & 59334.8 & 2021-04-30 & 1800 & 4096 & 122 & Tracking \\
\arrayrulecolor[gray]{0.6}\cline{2-7}\arrayrulecolor{black}
 & 59872.5 & 2022-10-20 & 1800 & 8192 & 61 & Swift Calibration \\
 \arrayrulecolor[gray]{0.6}\cline{2-7}\arrayrulecolor{black}
 & 59927.3 & 2022-12-14 & 1800 & 8192 & 61 & Swift Calibration \\
  \arrayrulecolor[gray]{0.6}\cline{2-7}\arrayrulecolor{black}
 & 59989.0 & 2023-02-13 & 1800 & 8192 & 61 & Swift Calibration \\ \hline
\multirow{5}{*}{B2021+51} & 59839.5 & 2022-09-17 & 5280 & 8192 & 61 & Swift Calibration \\
\arrayrulecolor[gray]{0.6}\cline{2-7}\arrayrulecolor{black}
 & 59900.4 & 2022-11-17 & 3600 & 8192 & 61 & Swift Calibration \\
 \arrayrulecolor[gray]{0.6}\cline{2-7}\arrayrulecolor{black}
 & 59961.3 & 2023-01-17 & 3600 & 8192 & 61 & Swift Calibration \\
 \arrayrulecolor[gray]{0.6}\cline{2-7}\arrayrulecolor{black}
 & 60024.1 & 2023-03-21 & 3600 & 8192 & 61 & Swift Calibration \\
 \arrayrulecolor[gray]{0.6}\cline{2-7}\arrayrulecolor{black}
 & 60083.9 & 2023-05-19 & 3600 & 8192 & 61 & Swift Calibration \\
\enddata
\end{deluxetable*}

\subsection{Dynamic Spectra}
\label{subsec:Dynamic Spectra}

We used DSPSR \footnote{\url{https://psrchive.sourceforge.net/}} \citep{van_Straten_2011} to fold the data in each frequency channel, 
\nc{with a sub-integration time chosen to be the closest value to 5 s that contains an integer number of pulse periods,} and then summed the horizontal and vertical polarizations to obtain the total intensity. Next, we computed the average pulse profile in order to determine the on-pulse and off-pulse regions. The region below the median of the profile was identified as the off-pulse region, and the region above 10\% of the peak amplitude with respect to the off-level was marked as the on-pulse region.
For each sub-integration and frequency channel, we calculated the mean intensity in the on-pulse region minus the mean intensity in the off-pulse region. The result of this calculation is the dynamic spectrum.

We used the level of the off-pulse signal to identify the radio-frequency interference (RFI). \nc{After experimentation, we set the RFI threshold as: threshold = min + 1.5 × (median - min), where min and median are calculated from the off-pulse signals across all pixels. Any pixel exceeding this threshold is flagged as RFI.}

Next, we applied Savitzky-Golay filters \citep{1964AnaCh..36.1627S} to smooth the dynamic spectra. Finally, we applied Wiener filters \citep{Wiener_1949, 2021MNRAS.506.2824L} to the dynamic spectra, filling the masked regions with the expected signal values obtained from the filter.

However, some signals in the unmasked regions still exhibited significant RFI. Therefore, we performed a second round of masking on the dynamic spectra to further remove RFI. We calculated the difference between the signal value in each unmasked pixel and its expected value from the Wiener filter and then computed the mean and standard deviation of these differences. If the difference for a given pixel deviated from the mean by more than three times the standard deviation, that pixel was additionally masked and its value filled by linear interpolation.

This second masking process removed only about 2–3\% of the dynamic spectrum data, but significantly improved the quality of the secondary spectrum for some observations. \nc{An example comparing the results before and after this process is shown in Figure \ref{fig:B1237+25_20200117_compare}.}

\subsection{Secondary Spectra}
\label{subsec:Secondary Spectra}

We obtained the secondary spectra by performing a fast Fourier transform (FFT) on the dynamic spectra and taking the square of the modulus of the results. Before performing the FFT, we interpolated each dynamic spectrum onto a grid of equal wavelength spacing, in order to eliminate the dependence of scintillation arc curvature on observing frequency (e.g., \citealt{Fallows_2014}; \citealt{2020ApJ...904..104R}; \citealt{2024MNRAS.527.7568O}). \nc{Note that this method smears discrete features along the arc, making it unsuitable for studies of discrete features such as inverted arclets \citep{theta_theta}. However, this does not hinder the detection and fitting of arcs.} The frequency-independent arc curvature $\eta$, such that $f_{\lambda}$ = $\eta f_{t}^{2}$ (where $f_{\lambda}$ is the differential time delay of wavelength conjugation), is given by

\begin{equation}
\eta = \frac{d_{\text{psr}} s(1 - s)}{2 v_{\text{sr}}^2}.
\label{eta_lambda}
\end{equation}

\noindent After this resampling, a Hanning window is typically applied to the outer 10\% of the dynamic spectra to ensure the edges are continuous, thereby reducing edge effects (e.g., \citealt{2020ApJ...904..104R}; \citealt{2024MNRAS.527.7568O}). However, this approach weakens the signals at the edges of the dynamic spectra. Therefore, we used reflection padding on the dynamic spectra, reflecting them across both their time and frequency axes.

Let us define the dynamic spectrum to be padded as $A(i,j)$ of size $m \times n$. After reflection padding, we obtain a padded spectrum $P(u,v)$ of size $(2m-2)\times(2n-2)$. Our padding method is expressed by the following equations:

\begin{equation}
P(u, v) = A\bigl(i(u),\, j(v)\bigr),\ (u \in [1,2m-2],\ v \in [1,2n-2]),
\end{equation}

\noindent where

\begin{equation}
\begin{aligned}
i(u) = \begin{cases}
u & (u \le m) \\
2m - u & (u > m)
\end{cases} \ \ , \\
\text{and} \quad j(v) = \begin{cases}
v & (v \le n) \\
2n - v & (v > n)
\end{cases} \ \ .
\end{aligned}
\end{equation}

\nc{This approach retains more of the signal at the cost of mixing signals of different parities. The secondary spectra obtained through this method are symmetric about $f_{t} = 0$ and cannot preserve asymmetries in the data, potentially obscuring some arcs when multiple fine arcs are present. Through careful comparison of secondary spectra obtained with and without this method, we find that none of the arcs in our study are obscured. Note that this method is not suitable for fitting parabolic arcs with apexes offset from the origin or for identifying arc substructures. The parabolic arcs in our study show no detectable apex offset (see APPENDIX \ref{Testing for arc apex offsets} for detailed discussion). This method reduces the background noise in the secondary spectra, as shown in Figure \ref{fig:B1237+25_20200117_compare}.}

\subsection{Scintillation Arc Search and Curvature Measurement}
\label{subsec:Arc Curvature}

We rescaled the secondary spectrum intensity on a logarithmic scale, using dB as the unit of intensity. We set the noise floor to 0 dB, and any data below the noise floor was excluded from the scintillation arc search and curvature calculation.

To search for \nc{scintillation arcs}, we computed the mean intensity of the secondary spectrum along parabolic arc paths. For low arc curvatures (below 100 $\mathrm{m}^{-1}\ \mathrm{mHz}^{-2}$), the mean intensity was calculated out to $f_\lambda$ = 5000 $\mathrm{m}^{-1}$. We identified the peak of the resulting mean-intensity curve as the signal associated with the scintillation arc. This process is known as the generalized Hough transform \citep{BALLARD1981111, Bhat_2016}. The generalized Hough transform and its variation (e.g., \citealt{Xu_2018}; \citealt{2020ApJ...904..104R}) are common methods for searching and measuring scintillation arcs. \nc{Before identifying the arc signal, we reduce noise using an FFT-based low-pass filter to smooth the intensity curve along each parabolic path. The cutoff frequency was manually optimized for each pulsar and arc based on the noise characteristics.}

The method for determining the scintillation arc curvature is as follows. First, we measured the noise intensity in the secondary spectrum far from the arc region. Next, we took the position of the peak of the smoothed intensity curve as the best-fit curvature value. Then, we fitted a Gaussian curve to data points with an intensity greater than the peak intensity minus one noise level. The standard deviation $\sigma$ of the fitted Gaussian represents the uncertainty in the arc curvature. This method is similar to those used by e.g., \cite{2020ApJ...904..104R}, \cite{McKee_2022}, and \cite{2024MNRAS.527.7568O}.

\renewcommand{\arraystretch}{1.5}
\setlength{\tabcolsep}{1.8pt}
\begin{deluxetable*}{ccccccccccc}
\tablefontsize{\footnotesize}
\tablecaption{The curvature of scintillation arcs observed for PSR B1237+25 across multiple observations (Units: $\mathrm{m}^{-1}\,\mathrm{mHz}^{-2}$).}
\label{tab:B1237+25_eta}
\tablewidth{0pt}
\tablehead{
\colhead{} & \multicolumn{10}{c}{MJD and UTC Date (y-m-d)} \\
\cline{2-11}
\colhead{Arc ID} & \colhead{\begin{tabular}{c}58808.1 \\[-5pt] 2019-11-21\end{tabular}} & \colhead{\begin{tabular}{c}58865.9 \\[-5pt] 2020-01-17\end{tabular}} & \colhead{\begin{tabular}{c}59231.8 \\[-5pt] 2021-01-17\end{tabular}} & \colhead{\begin{tabular}{c}59297.7 \\[-5pt] 2021-03-24\end{tabular}} & \colhead{\begin{tabular}{c}59837.2 \\[-5pt] 2022-09-15\end{tabular}} & \colhead{\begin{tabular}{c}59899.1 \\[-5pt] 2022-11-16\end{tabular}} & \colhead{\begin{tabular}{c}59959.0 \\[-5pt] 2023-01-14\end{tabular}} & \colhead{\begin{tabular}{c}60021.6 \\[-5pt] 2023-03-18\end{tabular}} & \colhead{\begin{tabular}{c}60090.5 \\[-5pt] 2023-05-26\end{tabular}} & \colhead{\begin{tabular}{c}60143.4 \\[-5pt] 2023-07-18\end{tabular}}
}
\startdata
A & 2.62±0.09 & 2.58±0.07 & 2.54±0.13 & 2.62±0.06 & 2.58±0.08 & 2.58±0.07 & 2.69±0.11 & 2.58±0.08 & 2.58±0.05 & 2.61±0.10 \\
\arrayrulecolor[gray]{0.6}\cline{1-11}\arrayrulecolor{black}
B & - & 4.28±0.12 & - & - & - & - & 4.05±0.27 & - & - & - \\
\arrayrulecolor[gray]{0.6}\cline{1-11}\arrayrulecolor{black}
C & - & 7.14±0.21 & - & - & - & - & - & - & 7.09±0.10 & - \\
\arrayrulecolor[gray]{0.6}\cline{1-11}\arrayrulecolor{black}
D & 8.63±0.20 & 8.87±0.16 & 9.51±0.35 & 9.09±0.37 & 9.01±0.35 & 8.56±0.26 & 8.95±0.23 & 9.29±0.33 & 8.82±0.11 & 8.70±0.23 \\
\arrayrulecolor[gray]{0.6}\cline{1-11}\arrayrulecolor{black}
E & - & 30.7±2.4 & - & - & - & - & 27.1±3.9 & - & - & - \\
\arrayrulecolor[gray]{0.6}\cline{1-11}\arrayrulecolor{black}
F & - & 65.3±6.5 & - & 68.4±6.0 & - & 60.3±2.8 & 65.0±7.8 & 65.0±4.3 & 61.3±1.8 & 60.0±3.8 \\
\arrayrulecolor[gray]{0.6}\cline{1-11}\arrayrulecolor{black}
G & 120.8±3.4 & 158.4±3.0 & 158.2±6.8 & 179.5±5.9 & 104.4±3.4 & 116.4±2.6 & 156.6±4.4 & 182.4±8.2 & 150.2±3.5 & 114.8±3.7 \\
\arrayrulecolor[gray]{0.6}\cline{1-11}\arrayrulecolor{black}
H & 316±49 & 387±39 & 437±106 & - & - & 329±51 & 366±138 & - & 344±48 & 239±29 \\
\arrayrulecolor[gray]{0.6}\cline{1-11}\arrayrulecolor{black}
I & 778±178 & - & - & - & - & 734±104 & - & - & 670±216 & 461±67 \\
\arrayrulecolor[gray]{0.6}\cline{1-11}\arrayrulecolor{black}
J & - & 1465±154 & 1512±348 & - & 888±171 & - & 1767±803 & 3494±788 & 3488±829 & 1613±366 \\
\enddata
\end{deluxetable*}

\renewcommand{\arraystretch}{1.5}

\setlength{\tabcolsep}{10pt}
\begin{deluxetable*}{ccccc}
\tablecaption{The curvature of the single scintillation arc observed for PSR B1842+14 across multiple observations.}
\label{tab:B1842+14_eta}
\tablewidth{0pt}
\tablehead{
\colhead{\begin{tabular}{c}MJD \\[-5pt] UTC Date (y-m-d)\end{tabular}} & \colhead{\begin{tabular}{c}59334.8 \\[-5pt] 2021-04-30\end{tabular}} & \colhead{\begin{tabular}{c}59872.5 \\[-5pt] 2022-10-20\end{tabular}} & \colhead{\begin{tabular}{c}59927.3 \\[-5pt] 2022-12-14\end{tabular}} & \colhead{\begin{tabular}{c}59989.0 \\[-5pt] 2023-02-13\end{tabular}}
}
\startdata
Curvature ($\mathrm{m}^{-1}\,\mathrm{mHz}^{-2}$) & 2790$\pm$625 & 571$\pm$92 & 685$\pm$155 & 732$\pm$74 \\
\enddata
\end{deluxetable*}

\renewcommand{\arraystretch}{1.5}
\setlength{\tabcolsep}{10pt}
\begin{deluxetable*}{cccccc}
\tablecaption{The curvature of scintillation arcs observed for PSR B2021+51 across multiple observations (Units: $\mathrm{m}^{-1}\,\mathrm{mHz}^{-2}$).}
\label{tab:B2021+51_eta}
\tablewidth{0pt}
\tablehead{
\colhead{} & \multicolumn{5}{c}{MJD and UTC Date (y-m-d)} \\
\cline{2-6}
\colhead{Arc ID} & \colhead{\begin{tabular}{c}59839.5 \\[-5pt] 2022-09-17\end{tabular}} & \colhead{\begin{tabular}{c}59900.4 \\[-5pt] 2022-11-17\end{tabular}} & \colhead{\begin{tabular}{c}59961.3 \\[-5pt] 2023-01-17\end{tabular}} & \colhead{\begin{tabular}{c}60024.1 \\[-5pt] 2023-03-21\end{tabular}} & \colhead{\begin{tabular}{c}60083.9 \\[-5pt] 2023-05-19\end{tabular}}
}
\startdata
A & 84.8$\pm$4.4 & - & 78.7$\pm$1.8 & 78.9$\pm$3.1 & - \\
\arrayrulecolor[gray]{0.6}\cline{1-6}\arrayrulecolor{black}
B & 122.9$\pm$4.1 & 123.0$\pm$7.4 & 125.4$\pm$4.9 & - & - \\
\arrayrulecolor[gray]{0.6}\cline{1-6}\arrayrulecolor{black}
C & 221.1$\pm$12.5 & 221.4$\pm$9.3 & 231.8$\pm$14.3 & 241.1$\pm$11.0 & - \\
\arrayrulecolor[gray]{0.6}\cline{1-6}\arrayrulecolor{black}
D & 635$\pm$23 & 607$\pm$21 & 710$\pm$39 & 901$\pm$27 & - \\
\arrayrulecolor[gray]{0.6}\cline{1-6}\arrayrulecolor{black}
E & 1871$\pm$78 & 1938$\pm$71 & - & 6697$\pm$256 & 4319$\pm$287 \\
\arrayrulecolor[gray]{0.6}\cline{1-6}\arrayrulecolor{black}
F (?) & 7224$\pm$872 & 14893$\pm$1719 & 2141$\pm$79 & - & - \\
\enddata
\tablecomments{The question mark in "Arc ID F (?)" denotes that the grouping of these observations is tentative. For arc group F, due to limited observation data and significant curvature variations, we cannot confirm whether these arcs originate from the same screen.}
\end{deluxetable*}

\begin{figure*}[ht]
\centering
\includegraphics[width=0.9\textwidth]{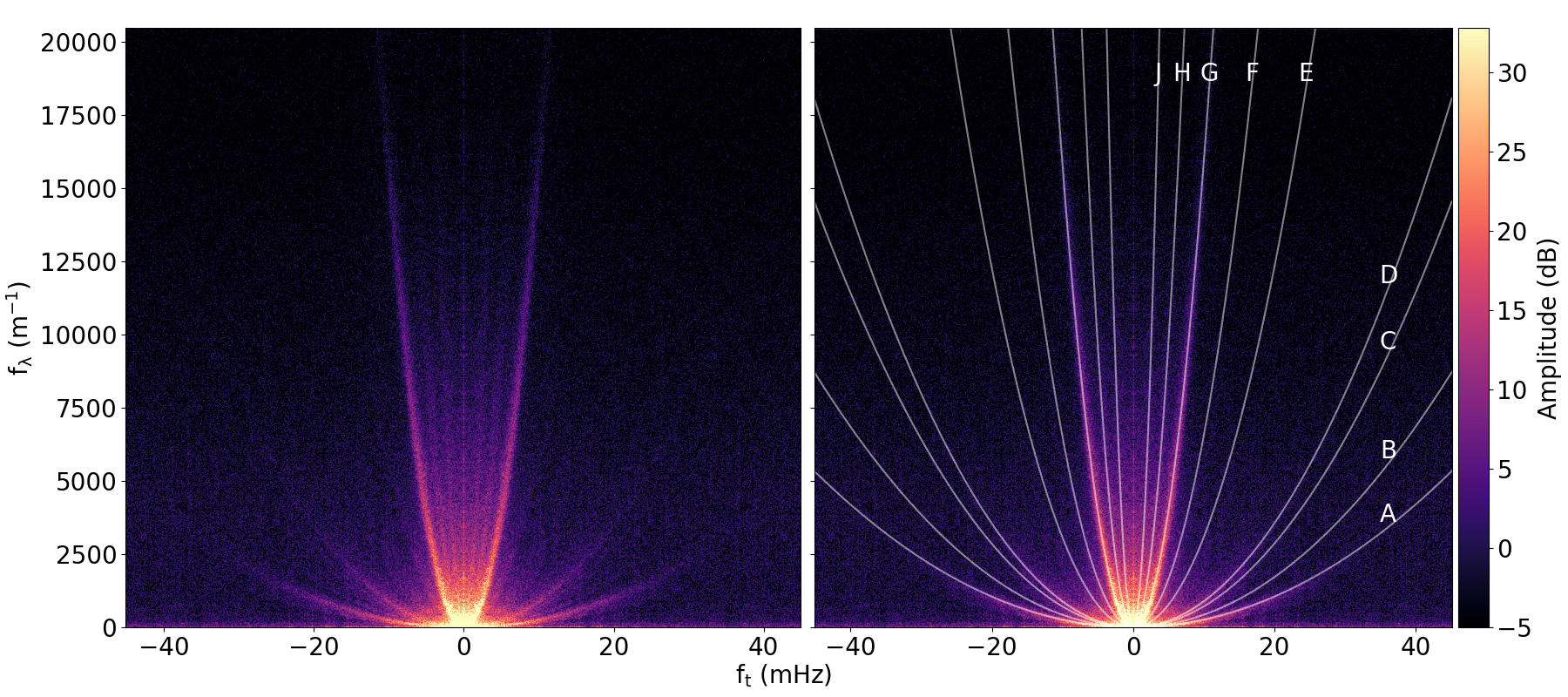}
\caption{Ten scintillation arcs were observed in pulsar B1237+25. Among them, nine scintillation arcs (excluding arc I) were identified during the observation on January 17, 2020. The left panel shows the secondary spectrum of that day, while the right panel shows the fitted scintillation arcs, labeled from A to J (excluding I) in order of increasing curvature.}
\label{fig:B1237+25_20200118_9_arc}
\end{figure*}

\begin{figure*}[ht]
\centering
\includegraphics[width=0.9\textwidth]{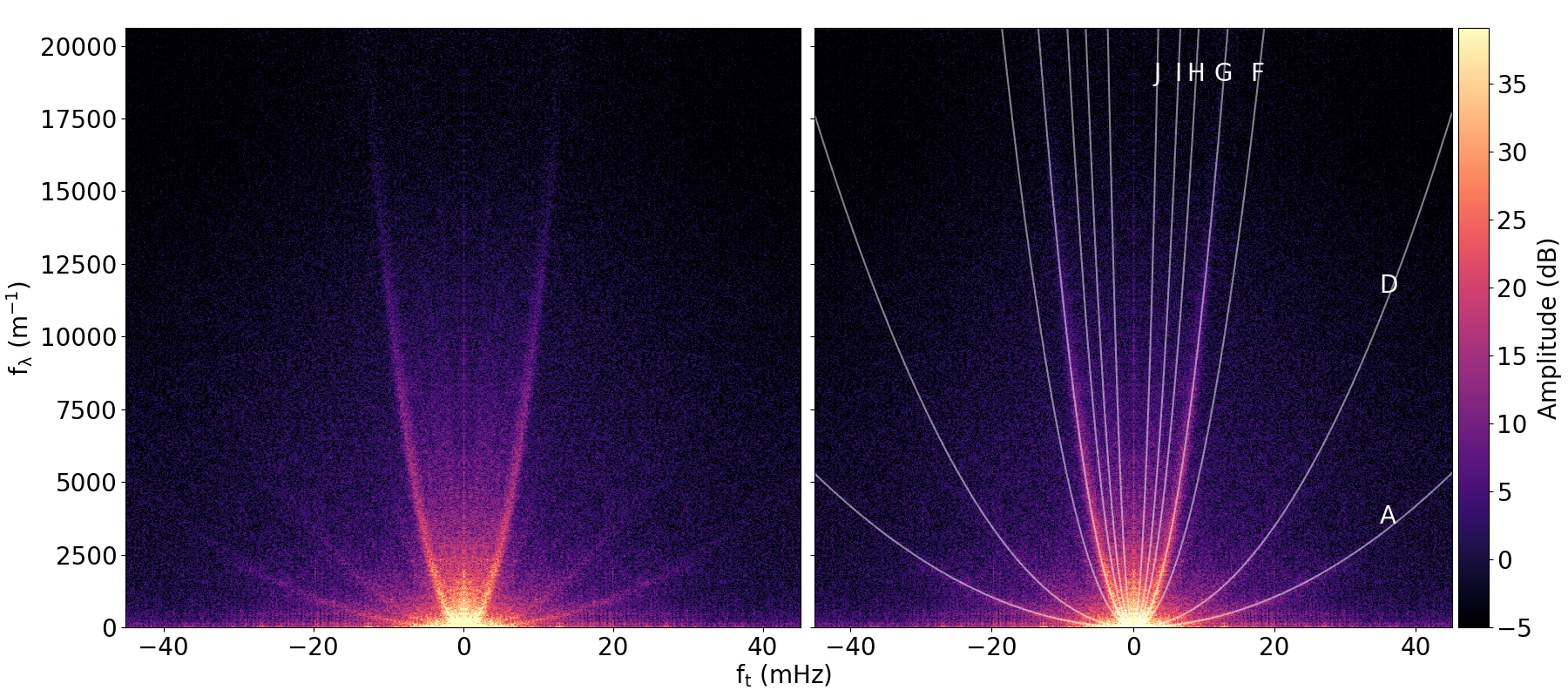}
\caption{Ten scintillation arcs were observed in pulsar B1237+25. Among them, seven scintillation arcs were identified during the observation on July 18, 2023. The left panel shows the secondary spectrum of that day, while the right panel shows the fitted scintillation arcs, labeled from A to J (excluding the unobserved arcs B, C, and E) in order of increasing curvature.}
\label{fig:B1237+25_20230718_7_arc}
\end{figure*}

\begin{figure*}[ht]
\centering
\includegraphics[width=0.9\textwidth]{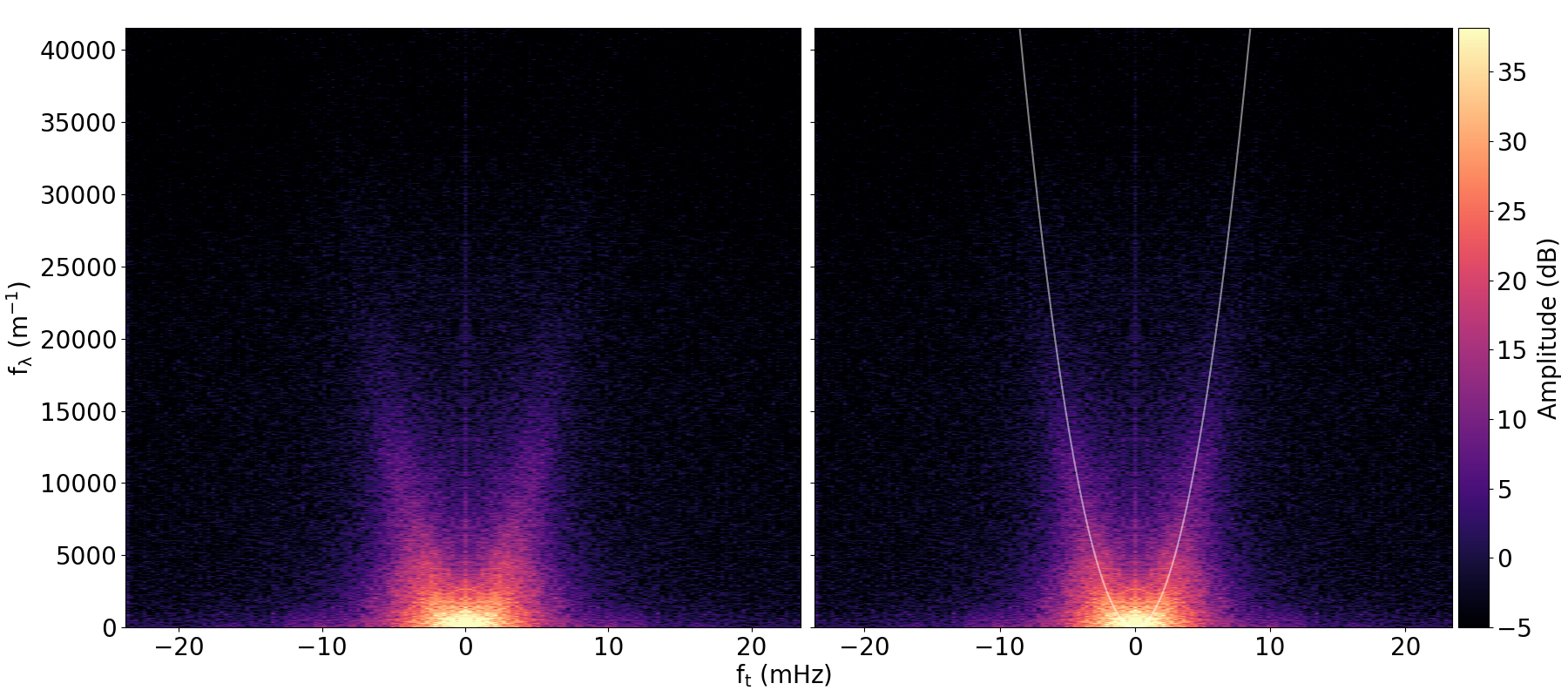}
\caption{One scintillation arc was identified in pulsar B1842+14 across four observations. This figure shows the secondary spectrum obtained from the observation on October 20, 2022, with the fitted scintillation arc shown in the right panel.}
\label{fig:B1842+14_20221020_all_arc}
\end{figure*}

\begin{figure*}[ht]
\centering
\includegraphics[width=0.9\textwidth]{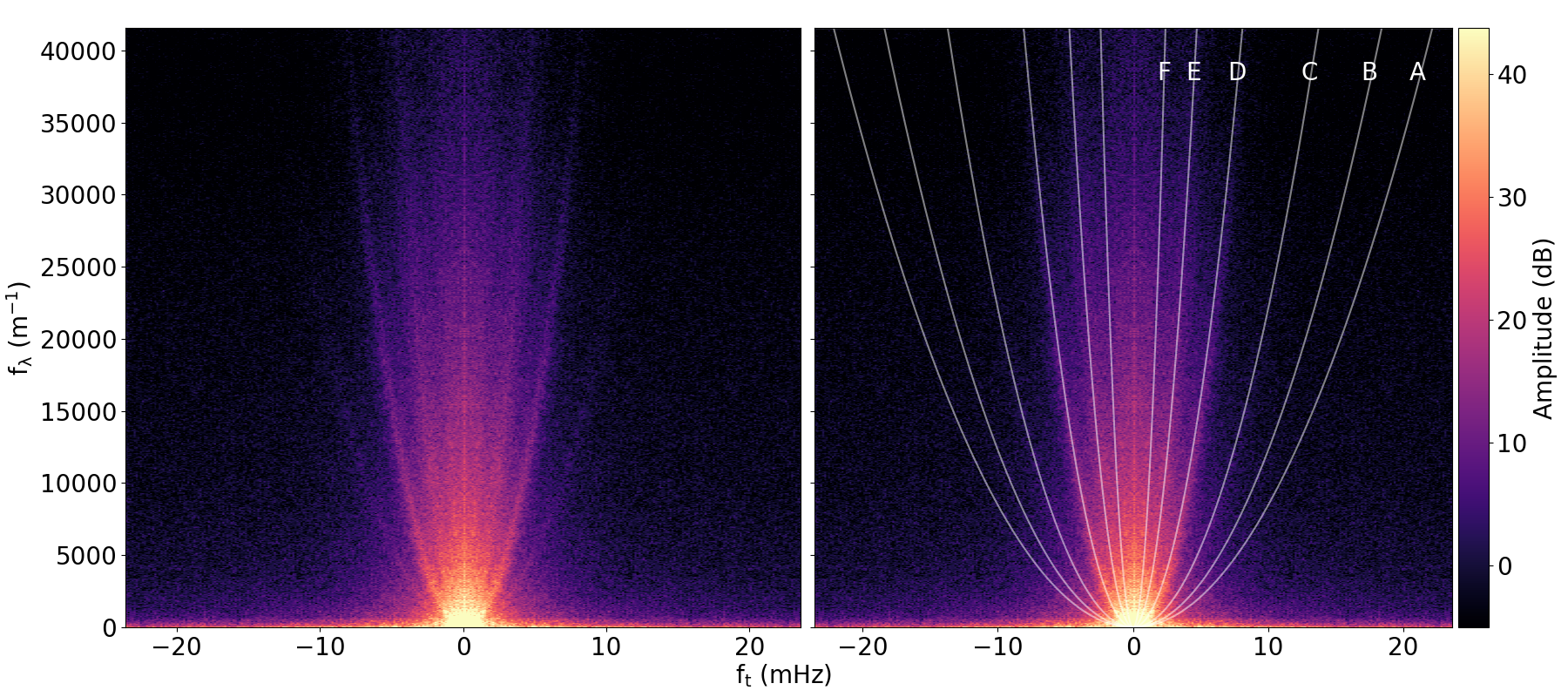}
\caption{The secondary spectrum of pulsar B2021+51 was obtained from the observation on September 17, 2022, with six scintillation arcs identified. The right panel shows the fitted scintillation arcs, labeled from A to F in order of increasing curvature.}
\label{fig:B2021+51_20220917_all_arc}
\end{figure*}

\section{Results} \label{results}

We observed scintillation arcs in all observations (as shown in Table \ref{tab:B1237+25_eta}, \ref{tab:B1842+14_eta}, and \ref{tab:B2021+51_eta}). The observed scintillation arcs varied in structure, with some being more diffuse (e.g., PSR B1842+14, as shown in Figure \ref{fig:B1842+14_20221020_all_arc}) and others being finer (e.g., the scintillation arc G of PSR B1237+25, as shown in Figure \ref{fig:B1237+25_20200118_9_arc} and \ref{fig:B1237+25_20230718_7_arc}). We note that except for arc F of pulsar B2021+51, all other scintillation arcs exhibited sharp parabolic arc characteristics, consistent with strongly anisotropic scintillation screens. For each anisotropic arc, we modeled its annual modulation using the Markov Chain Monte Carlo (MCMC) method to calculate the fractional distance $s$ of the scintillation screen, the velocity component of the scintillation screen along the scattering direction $v_{{\text{screen},~\zeta}}$, and the scattering direction \(\mathbf{n}_{\text{screen}}\), as shown in Table \ref{tab:parameters_screens}. We express the measured direction of the scattering direction using the azimuth angle \(\zeta\) as degrees east of north. The MCMC sampling was performed using the Python package emcee \citep{2013PASP..125..306F}, which efficiently explores the parameter space to generate posterior distributions for the parameters. 

However, we were unable to calculate all three parameters for every scintillation screen when modeling the modulation. As discussed in Section \ref{sec:Theory of annul modulation}, in cases of anisotropic scattering, the direction of the scattering direction of the scintillation screen is the most easily constrained parameter, as it only requires knowledge of the point of minimum curvature. The fractional distance $s$ is the next most easily constrained, while the screen's projected velocity $v_{{\text{screen},~\zeta}}$ is the most difficult to constrain. For some screens, we could only calculate partial parameters based on certain assumptions. Below are the situations we encountered and our approaches, respectively:

1. \textbf{Substantial errors}. The relative errors in the curvature measurements of the scintillation arc H for PSR B1237+25 are substantial, resulting in poor constraints on the screen parameters. In this case, considering that the projected velocity $v_{{\text{screen},~\zeta}}$ of the screen is often relatively small and is the most difficult to constrain, we assumed its value to be 0 and, after modeling, could only obtain estimates of $s$ and \(\zeta\).

2. \textbf{Weak modulation}. When the scintillation screen is close to the pulsar, i.e., \(s{\mathbf{v}_{\text{Earth}}}\cdot \mathbf{n}_{\text{screen}} \ll (1-s)  \mathbf{v}_{\text{psr}}\cdot \mathbf{n}_{\text{screen}} \), the modulation amplitude of the arc curvature is not significantly greater than the measurement error. For example, the difference between the maximum and minimum curvature of scintillation arc A for PSR B1237+25 is only 0.15 \(\mathrm{m}^{-1}\,\mathrm{mHz}^{-2}\), while the measurement error ranges from 0.05 \(\mathrm{m}^{-1}\,\mathrm{mHz}^{-2}\) to 0.13 \(\mathrm{m}^{-1}\,\mathrm{mHz}^{-2}\) (see Table \ref{tab:B1237+25_eta}). Scintillation arcs D and F of PSR B1237+25 (see Table \ref{tab:B1237+25_eta}) and scintillation arcs A, B, and C of PSR B2021+51 (see Table \ref{tab:B2021+51_eta}) also exhibit this characteristic. In such cases with small modulation amplitude, we could only constrain one screen parameter. Considering that $v_{{\text{screen},~\zeta}}$ is often relatively small, and these scintillation screens are relatively close to the pulsar, satisfying \( v_{{\text{screen},~\zeta}} \ll (1-s) \mathbf{v}_{\text{psr}}\cdot \mathbf{n}_{\text{screen}} \), we assumed $v_{{\text{screen},~\zeta}}$ to be 0. We also assumed the main scattering direction $\mathbf{n}_{\text{screen}}$ is aligned with the effective velocity (see e.g. \citealt{theta_theta}),

\begin{eqnarray}
\mathbf{v}_{\mathrm{eff}} \equiv-\frac{1}{s} \mathbf{v}_{\mathrm{screen}}+\mathbf{v}_{\mathrm{Earth}}+\frac{1-s}{s} \mathbf{v}_{\mathrm{psr}}.
\label{eq:V_eff}
\end{eqnarray}

\noindent After fitting, the resulting $s$ value given under these assumptions represents its expected maximum value, which is the minimum distance between the scintillation screen and Earth.

3. \textbf{Insufficient number of detections}. Some scintillation arcs could only be identified in observations conducted on dates with close day-of-year positions. Scintillation arcs B and E of PSR B1237+25 (see Table \ref{tab:B1237+25_eta}) were only observed on January 17, 2020, and January 14, 2023, which is effectively equivalent to having only a single observation. In such cases, we assumed $v_{{\text{screen},~\zeta}}$ to be 0 and $\mathbf{n}_{\text{screen}}$ to be aligned with $\mathbf{v}_{\text{eff}}$ to calculate the value of $s$. We obtained two solutions for $s$: the maximum value of $s$ if the scintillation screen is close to the pulsar, and the minimum value of $s$ if the scintillation screen is close to Earth.

4. \textbf{Insufficient sampling of the modulation curve}. For scintillation arc I of PSR B1237+25, two of the four observations were conducted on dates with close day-of-year positions (November 21, 2019, and November 16, 2022), lacking sufficient phase information about the annual modulation of curvature. This made constraining the screen parameters difficult. In this case, we assumed $v_{{\text{screen},~\zeta}}$ to be 0, and obtained estimates of $s$ and \(\zeta\). For PSR B1237+25, the scintillation arc C was observed twice with about a 129-day interval in day-of-year positions, and their measured curvature values were very close, even less than the measurement error. Arc C could be a weakly modulated arc, or it might be a strongly modulated arc that happened to be observed on two dates with very similar curvature values. Therefore, we assumed $v_{{\text{screen},~\zeta}}$ to be 0 and $\mathbf{n}_{\text{screen}}$ to be aligned with $\mathbf{v}_{\text{eff}}$ to calculate the value of s. We obtained two solutions for s.
The scintillation arc F of PSR B2021+51 had fewer observation data points and large differences in measured curvature. We could not determine whether this group of scintillation arcs originated from the same screen, so we did not fit it.

\renewcommand{\arraystretch}{1.7}
\setlength{\tabcolsep}{6pt}
\begin{deluxetable*}{c|ccccc}
\tablefontsize{\footnotesize}
\tablecaption{Parameters of scintillation screens observed in the directions of pulsars B1237+25, B1842+14, and B2021+51. The parameter $s$ represents the fractional distance of the scintillation screen, $d_\text{screen}$ represents the distance from the scintillation screen to the Earth, $\zeta$ represents the azimuth angle (degrees east of north) of the scattering direction, and $v_{{\text{screen},~\zeta}}$ represents the velocity component along the scattering direction. For screens without a $\zeta$ value, we assume their scattering direction and effective velocity are aligned. For screens without a $v_{{\text{screen},~\zeta}}$ value, we assume that their $v_{{\text{screen},~\zeta}}$ is zero.}
\label{tab:parameters_screens}
\tablewidth{0pt}
\tablehead{
\colhead{Pulsar} & \colhead{Screen ID} & \colhead{$s$} & \colhead{$d_\text{screen}$ ($\text{pc}$)} & \colhead{$\zeta$ ($^{\circ}$)} & \colhead{$v_{{\text{screen},~\zeta}}$ (km/s)}
}
\startdata
\multirow{13}{*}{B1237+25} & A & $0.0435_{-0.0027}^{+0.0031}$ & $824^{+61}_{-53}$ & - & - \\
\arrayrulecolor[gray]{0.6}\cline{2-6}\arrayrulecolor{black}
~ & B & \begin{tabular}{c}$0.0691^{+0.0051}_{-0.0045}$ \\[-5pt] $0.999888^{+8 \times 10^{-6}}_{-8 \times 10^{-6}}$\end{tabular} & \begin{tabular}{c} $802^{+60}_{-52}$ \\[-5pt] $0.0969^{+0.0103}_{-0.0093}$ \end{tabular} & - & - \\
\arrayrulecolor[gray]{0.6}\cline{2-6}\arrayrulecolor{black}
~ & C & \begin{tabular}{c}$0.110^{+0.007}_{-0.007}$ \\[-5pt] $0.999837^{+1.1 \times 10^{-5}}_{-1.1 \times 10^{-5}}$ \end{tabular}& \begin{tabular}{c}$767^{+57}_{-50}$ \\[-5pt] $0.140^{+0.014}_{-0.013}$ \end{tabular}& - & - \\
\arrayrulecolor[gray]{0.6}\cline{2-6}\arrayrulecolor{black}
~ & D & $0.134^{+0.009}_{-0.008}$ & $746^{+56}_{-49}$ & - & - \\
\arrayrulecolor[gray]{0.6}\cline{2-6}\arrayrulecolor{black}
~ & E & \begin{tabular}{c}$0.338^{+0.022}_{-0.021}$ \\[-5pt] $0.999227^{+7.1 \times 10^{-5}}_{-7.7 \times 10^{-5}}$\end{tabular} & \begin{tabular}{c} $570^{+46}_{-41}$ \\[-5pt] $0.667^{+0.083}_{-0.075}$ \end{tabular} & - & - \\
\arrayrulecolor[gray]{0.6}\cline{2-6}\arrayrulecolor{black}
~ & F & $0.512^{+0.018}_{-0.017}$ & $421^{+35}_{-31}$ & - & - \\
\arrayrulecolor[gray]{0.6}\cline{2-6}\arrayrulecolor{black}
~ & G & $0.691^{+0.026}_{-0.028}$ & $267^{+32}_{-28}$ & $124.7^{+5.1}_{-5.2}$ & $-1.2^{+8.6}_{-9.5}$ \\
\arrayrulecolor[gray]{0.6}\cline{2-6}\arrayrulecolor{black}
~ & H & $0.632^{+0.096}_{-0.151}$ & $321^{+130}_{-88}$ & $171.1^{+10.0}_{-10.5}$ & - \\
\arrayrulecolor[gray]{0.6}\cline{2-6}\arrayrulecolor{black}
~ & I & $0.756^{+0.114}_{-0.224}$ & $214^{+191}_{-102}$ & $175.1^{+11.6}_{-12.7}$ & - \\
\arrayrulecolor[gray]{0.6}\cline{2-6}\arrayrulecolor{black}
~ & J & $0.723^{+0.069}_{-0.099}$ & $241^{+86}_{-63}$ & $63.7^{+8.4}_{-7.3}$ & $-41^{+22}_{-37}$ \\
\hline
B1842+14 & - & $0.85^{+0.05}_{-0.10}$ & $2.4^{+2.1}_{-1.2}$ $\times 10^{2}$ & $122^{+11}_{-12}$ & $19^{+11}_{-11}$ \\
\hline
\multirow{5}{*}{B2021+51} & A & $0.0368^{+0.0062}_{-0.0047}$ & $1.93^{+0.31}_{-0.24}$ $\times 10^{3}$ & - & - \\
\arrayrulecolor[gray]{0.6}\cline{2-6}\arrayrulecolor{black}
~ & B & $0.0567^{+0.0093}_{-0.0072}$ & $1.88^{+0.31}_{-0.23}$ $\times 10^{3}$ & - & - \\
\arrayrulecolor[gray]{0.6}\cline{2-6}\arrayrulecolor{black}
~ & C & $0.100^{+0.015}_{-0.012}$ & $1.80^{+0.29}_{-0.22}$ $\times 10^{3}$ & - & - \\
\arrayrulecolor[gray]{0.6}\cline{2-6}\arrayrulecolor{black}
~ & D & $0.406^{+0.072}_{-0.069}$ & $1.19^{+0.24}_{-0.20}$ $\times 10^{3}$ & $172.4^{+10.9}_{-9.2}$ & $32.7^{+8.1}_{-9.1}$ \\
\arrayrulecolor[gray]{0.6}\cline{2-6}\arrayrulecolor{black}
~ & E & $0.556^{+0.041}_{-0.038}$ & $887^{+167}_{-132}$ & $20.3^{+4.3}_{-4.3}$ & $-12.1^{+3.7}_{-3.8}$ \\
\enddata
\end{deluxetable*}

\begin{figure*}[ht]
\centering
\includegraphics[width=\textwidth]{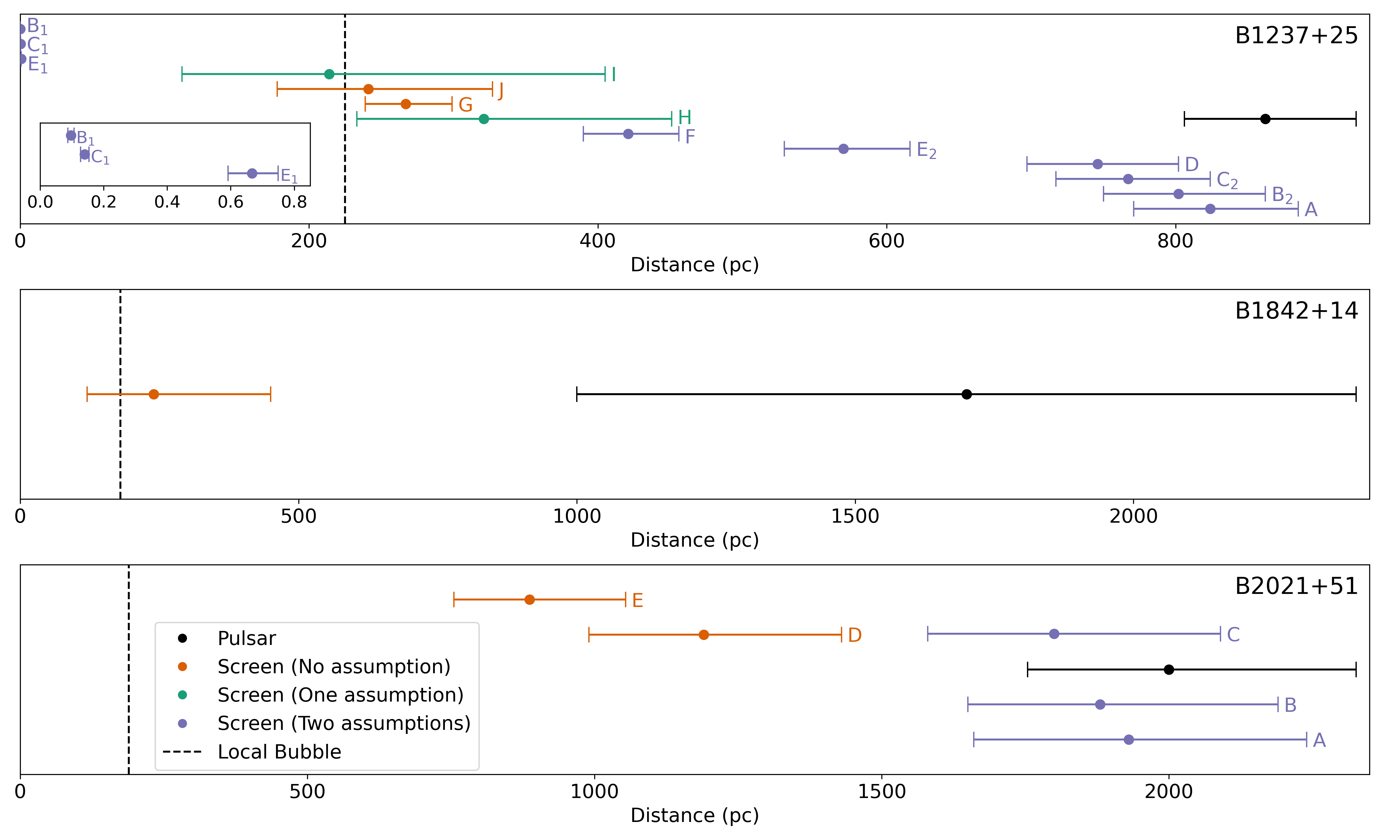}
\caption{Screen distance estimates for B1237+25 (top), B1842+14 (middle), and B2021+51 (bottom). The scintillation screen labeled ``No assumption'' indicates their positions were precisely calculated with all three parameters determined. The scintillation screen labeled ``One assumption'' indicates their positions were calculated based on one assumption that the velocity is zero. The scintillation screen labeled ``Two assumptions'' indicates their positions were calculated based on two assumptions: zero velocity and scattering direction aligned with the effective velocity. For pulsars with multiple scintillation screens, each screen has been marked with its ID. For PSR B1237+25, scintillation screens B, C, and E have two sets of distance solutions. For solutions close to the Sun, we label them as $\text{B}_1$, $\text{C}_1$, and $\text{E}_1$, while for solutions close to the pulsar, we label them as $\text{B}_2$, $\text{C}_2$, and $\text{E}_2$. The black dashed line represents the distance of the Local Bubble inner surface \citep{LB_2020, LB_2022}. \nc{The distance uncertainties of the screens include contributions from the pulsar distance uncertainty. For screens near the pulsar, this correlation causes their error bars to overlap with the pulsar distance uncertainties. The fractional screen distances $s$ are provided in Table \ref{tab:parameters_screens}.}}
\label{fig:pulsar_screens_1D}
\end{figure*}

\begin{figure*}[ht]
\centering
\includegraphics[width=0.75\textwidth]{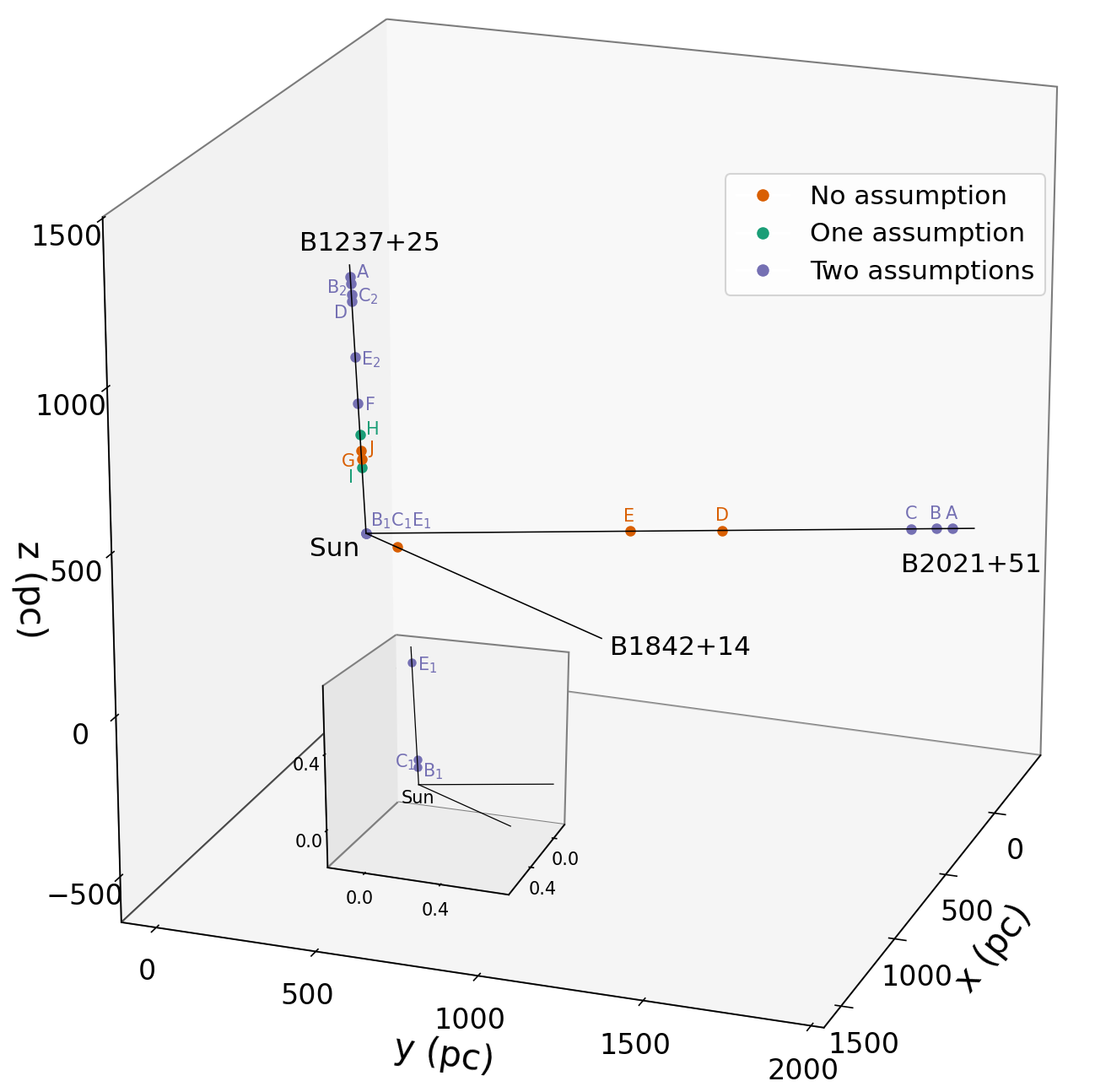}
\caption{The heliocentric Galactic Cartesian coordinates of three pulsars (PSR B1237+25, PSR B1842+14, and PSR B2021+51) and the scintillation screens identified in their directions are displayed in this figure. The scintillation screen labeled ``No assumption'' indicates their positions were precisely calculated with all three parameters determined. The scintillation screen labeled ``One assumption'' indicates their positions were calculated based on one assumption that the velocity is zero. The scintillation screen labeled ``Two assumptions'' indicates their positions were calculated based on two assumptions: zero velocity and scattering direction aligned with the effective velocity. For pulsars with multiple scintillation screens, each screen has been marked with its ID. For PSR B1237+25, scintillation screens B, C, and E have two sets of distance solutions. For solutions close to the Sun, we label them as $\text{B}_1$, $\text{C}_1$, and $\text{E}_1$, while for solutions close to the pulsar, we label them as $\text{B}_2$, $\text{C}_2$, and $\text{E}_2$. Black lines connect the position of the Sun (0, 0, 0) with the positions of the pulsars.}
\label{fig:screens_positions}
\end{figure*}

\subsection{PSR B1237+25}

We identified 10 scintillation arcs in the direction of pulsar B1237+25, as shown in Table \ref{tab:B1237+25_eta}. Among them, 9 scintillation arcs could be identified in the observation data from January 17, 2020, as shown in Figure \ref{fig:B1237+25_20200118_9_arc}. \nc{Previously, three scintillation arcs were detected by \cite{Main_2023_107P} using the MeerKAT Thousand Pulsar Array at 856 – 1712 MHz, one scintillation arc was detected by \cite{Fadeev_2018} using the GBT and Arecibo radio telescope at 316 – 332 MHz, who also estimated the screen distance, and only a diffuse blend of power in the secondary spectrum was observed by \cite{Wu_2022} using LOFAR at 120 – 180 MHz.}

Scintillation arc G had the highest brightness and was identified in all 10 observations. We measured and modeled the curvature modulation of arc G (as shown in Figure \ref{fig:B1237+25_G_eta_MCMC}). We calculated that the fractional distance $s$ of scintillation screen G is $0.691^{+0.026}_{-0.028}$ (corresponding to a distance of $267^{+32}_{-28}$ pc from Earth), the scattering direction \(\zeta\) is $124.7^{+5.1}_{-5.2}$ \(^\circ\), and the velocity of the scintillation screen along the scattering direction $v_{{\text{screen},~\zeta}}$ is \(-1.2^{+8.6}_{-9.5}\) km/s. \nc{\cite{Fadeev_2018} estimated the distance to the only scintillation screen they detected in this direction to be 0.23 $\pm$ 0.05 kpc, which is consistent with the distance to screen G}. We also modeled and measured the curvature modulation of scintillation arcs J and I, successfully calculating all three parameters.

The measurement error of the curvature of scintillation arc H was relatively large, making it difficult to constrain the scintillation screen parameters, especially the $v_{{\text{screen},~\zeta}}$. Therefore, we assumed $v_{{\text{screen},~\zeta}}$ to be 0 and only calculated the fractional distance $s$ and the scattering direction. Scintillation arc I was identified in only 4 observations, with two occurring on dates in close proximity, preventing us from obtaining sufficient phase information about the curvature modulation. Hence, we also assumed $v_{{\text{screen},~\zeta}}$ to be 0.

Scintillation arc A is bright, so it was identified in all 10 observations. However, its modulation intensity was low. This suggests that the fractional distance $s$ of the scintillation screen corresponding to arc A is close to zero, with almost no modulation from Earth's annual orbital motion. We could not decouple the parameters determining the curvature by modeling its modulation, so we assumed its velocity to be 0 and $\mathbf{n}_{\text{screen}}$ to be aligned with $\mathbf{v}_{\text{eff}}$ to calculate the distance of the corresponding scintillation screen. We calculated that the fractional distance $s$ of scintillation screen A is $0.0435_{-0.0027}^{+0.0031}$. This estimated fractional distance likely differs significantly from the actual position of the scintillation screen; it is merely an upper limit of the actual value. Similarly, scintillation arcs D and F also have low modulation intensity, so they were processed in the same way.

Scintillation arcs B and E were only identified in observations on November 21, 2019, and January 17, 2020. The two dates correspond to days of the year that are too close together, so we essentially have no annual modulation information for them. We applied the assumptions that $v_{{\text{screen},~\zeta}}$ is 0 and that $\mathbf{n}_{\text{screen}}$ is aligned with $\mathbf{v}_{\text{eff}}$ to estimate their fractional distance $s$. In this case, $s$ will have two solutions: a solution close to 0, which is the upper limit when the screen is close to the pulsar, and a solution close to 1, which is the lower limit when the screen is close to Earth. For arc C, we only have two observed values that are close together, giving us little annual modulation information, so it was processed in the same way.

\subsection{PSR B1842+14}

We conducted 4 observations of pulsar B1842+14, each identifying 1 scintillation arc, as shown in Table \ref{tab:B1842+14_eta}. The reason for observing only one arc may be its relatively low brightness, which is an order of magnitude lower than the other two pulsars studied in this paper (as shown in Table \ref{tab:PSRs}). \cite{2020RAA....20...76Y} observed 2 scintillation arcs from this pulsar on August 28, 2018, one with a curvature of 0.12 \(\pm\) 0.05 s${^3}$ at 1400 MHz, which is 785 \(\pm\) 327 $\mathrm{m}^{-1}\,\mathrm{mHz}^{-2}$. We compared this value with our results (as shown in Figure \ref{fig:B1842+14_eta_MCMC}). \cite{2020RAA....20...76Y} also observed another scintillation arc with a curvature of 0.0010 \(\pm\) 0.0003 s${^3}$ at 1400 MHz, which we did not observe. Possible reasons include that this arc has moved away from the line of sight to the pulsar, or it may be due to significant noise interference. Further research on this issue will require longer continuous observations to obtain data with higher signal-to-noise ratios. The scintillation arc we observed was relatively diffuse (as shown in Figure \ref{fig:B1842+14_20221020_all_arc}), resulting in larger errors in the curvature measurements. The limited dataset of 4 observations reduces the precision of our parameter-fitting results. We hope to conduct more observations of pulsar B1842+14 in the future to improve the modeling accuracy. We modeled the annual modulation of the scintillation arc (as shown in Figure \ref{fig:B1842+14_eta_MCMC}). Through modeling the annual modulation, we determined that the fractional distance $s$ of the scintillation screen is $0.85^{+0.05}_{-0.10}$, corresponding to a distance of $0.24^{+0.21}_{-0.12}$ kpc; the scattering direction $\zeta$ is $122^{+11}_{-12}$ $^\circ$; and the velocity component along the scattering direction $v_\text{scr}$ is $19^{+11}_{-11}$ km/s. \cite{2020RAA....20...76Y} reported a result of 0.3 $\pm$ 0.2 kpc for this arc based on a single observation assuming $\mathbf{n}_{\text{screen}}$ aligns with $\mathbf{v}_\text{eff}$ and $v_\text{screen}$ = 0. The large uncertainty for these screen distance measurements is mainly attributed to the uncertainty in the pulsar distance. Accurate pulsar distance measurements are vital for obtaining more precise screen distance determinations.

\subsection{PSR B2021+51}

We discovered at least 6 scintillation arcs in the observation data of pulsar B2021+51 (see Table \ref{tab:B2021+51_eta}), with 6 scintillation arcs identifiable in the observation data from September 17, 2022, as shown in Figure \ref{fig:B2021+51_20220917_all_arc}. \nc{Previously, \cite{Stinebring_2022} observed 2 scintillation arcs using the Green Bank Telescope.}

Among these, the scintillation arc E was the brightest. Our model fitting yielded the following three parameters for this scintillation screen: a fractional distance $s$ of $0.556^{+0.041}_{-0.038}$ (corresponding to a distance of \(887^{+167}_{-132}\) pc), a scattering direction of $20.3^{+4.3}_{-4.3}$ \(^{\circ}\), and a projected velocity of $-12.1^{+3.7}_{-3.8}$ km/s. For scintillation arc D, we also successfully fitted all three parameters of the scintillation screen.

Scintillation arcs A, B, and C all have low modulation intensity. We could not obtain all three parameters for these screens through modeling and could only estimate their fractional distances based on the assumption that the screen velocity is 0 and the scattering direction is aligned with the effective velocity.

The F group of scintillation arcs exhibited characteristics of an isotropic screen. However, there were relatively few observation data points (only 3), and the measurements varied greatly (see Table \ref{tab:B2021+51_eta}). We need more observation results to determine whether the arcs observed in these 3 instances originated from the same scintillation screen.

\nc{\subsection{Distribution and potential sources of scintillation screens}}

Figures \ref{fig:pulsar_screens_1D} and \ref{fig:screens_positions} summarize our screen detection and distance determination for all three pulsars. In total, we precisely determined the distances of 5 scintillation screens, calculated the distances of 2 scintillation screens using the single assumption that screen velocity equals zero, and calculated the distances of 9 scintillation screens using two assumptions: zero screen velocity and scattering direction aligned with the effective velocity.

We find that scintillation screens are distributed throughout the entire paths between Earth and the pulsars. This widespread distribution suggests that electron density fluctuations capable of causing pulsar scintillation exist across diverse regions of interstellar space.

PSR B1237+25, with a Galactic latitude of 86.5 degrees, exhibits four scintillation screens (screens G, H, I, and J) that appear at distances that roughly coincide with the Local Bubble boundary. The Local Bubble boundary is one of the known sources of pulsar scintillation \citep{Bhat_1998}. Therefore, one potential application of pulsar scintillation is to constrain the shape of the Local Bubble. Measurements of the Local Bubble's shape in high-latitude directions have larger errors \citep{Yeung_2024}, and studying scintillation screens towards high-latitude pulsars like the brightest scintillation arc (G) in the direction of PSR B1237+25 could potentially help us constrain the shape of the Local Bubble. To achieve this goal, more precise determinations of pulsar distances are necessary. A notable example is PSR B1842+14, which has significant distance uncertainty (1.7 $\pm$ 0.7 kpc), leading to large errors in the calculated distance of the screen in that direction ($2.4^{+2.1}_{-1.2} \times 10^{2}$ pc). PSR B1842+14 exhibits a single scintillation screen which also appears at distances that roughly coincide with the Local Bubble boundary.

Only three potential screens (alternative solutions for B, C, and E of PSR B1237+25) might reside within the Local Bubble, and these represent just one of two possible solutions for each screen, with the alternative placements being at much greater distances. These extremely nearby solutions, if physical, would be associated with very local interstellar structures at distances of only $<$ 1 pc. \nc{Scintillation screens close to Earth usually produce arcs with large annual curvature variation, depending on the pulsar velocity. When the annual curvature variation is large, it makes it relatively straightforward to resolve the distance degeneracy through observations at multiple epochs throughout the year.}

\nc{Screens E and D for PSR B2021+51 are located far from both the Local Bubble and the pulsar itself, and thus appear unrelated to the Local Bubble or pulsar local environments,} suggesting they originate from general ISM structures distributed along the line of sight.

\nc{PSR B1237+25 and PSR B2021+51 have close screens that may be associated with their local environments. However, the spin-down luminosities $\dot{E}$ of PSR B1237+25 and PSR B2021+51 are low (1.4 $\mathrm{\times 10^{31} ergs^{-1}}$ and 8.2 $\mathrm{\times 10^{32}ergs^{-1}}$, respectively), so they are unlikely to have observable bow shocks. If bow shocks were present, the bow shock radius along the line of sight would be $\lesssim$ O(1000)~AU for PSR B1237+25, and $\lesssim$ O(10000)~AU for PSR B2021+51 (more details in APPENDIX \ref{bow shock radii}). However, we only know that the closest screen to PSR B1237+25 is at a distance $\lesssim$~40 pc (see Table \ref{tab:parameters_screens}), while that for PSR B2021+51 is $\lesssim$~70 pc (see Table \ref{tab:parameters_screens}), so we cannot determine whether the screens are at the bow shocks. If the closest screens are at their respective bow shocks, the relative velocity between the pulsars and the screens in the main scattering direction ($v_{\mathrm{psr},\zeta} - v_{\mathrm{screen},\zeta}$) would be O(1) $\mathrm{km~s^{-1}}$ (
APPENDIX \ref{bow shock radii}).}

\section{Conclusion}
\label{Conclusion}

We analyzed observational data from the FAST telescope and, benefiting from the high sensitivity of FAST and our improved data processing methods, identified a total of 16 scintillation screens in the directions of pulsars B1237+25, B1842+14, and B2021+51.  Specifically, 10 screens were identified for B1237+25, 1 for B1842+14, and 5 for B2021+51.

For 5 of these screens, we successfully constrained all three screen parameters (distance, velocity, and scattering direction), while for the others, we provided distance estimates based on simplified assumptions. The errors in scintillation screen distances primarily originate from pulsar distance uncertainties, which are particularly evident in the direction of PSR B1842+14 where the estimated distance error is very large. High-precision pulsar distance measurements are especially important for accurate screen distance determinations.

Our findings reveal that these scintillation phenomena occur throughout various regions of interstellar space between Earth and the observed pulsars. This distribution pattern indicates that electron density fluctuations capable of causing such scintillation are widespread throughout interstellar space. Several screens in our sample appear at distances that coincide with the Local Bubble boundary, particularly the brightest scintillation arc in the direction of PSR B1237+25. When confirmed to be associated with the Local Bubble boundary, the distance measurement of such scintillation screens would provide a valuable constraint on the Local Bubble shape at high Galactic latitudes. \nc{For screens located in close proximity to PSR B1237+25 and PSR B2021+51, the distance uncertainties and the uncertainty about the presence of bow shocks prevent us from determining whether these screens are associated with the pulsars' local environments.}

This study provides a rich observational sample of scintillation screens, offering insight into the remarkably diverse structures present in the interstellar medium. The findings contribute valuable data that will enable future correlations between scintillation screens and interstellar structures, which will help advance our understanding of their astrophysical origins. Further observations with longer durations will enhance parameter precision, ultimately helping to elucidate the physical nature of these interstellar medium structures.


\section{Acknowledgments}
 
This work made use of the data from FAST (Five-hundred-meter Aperture Spherical radio Telescope).  FAST is a Chinese national mega-science facility, operated by National Astronomical Observatories, Chinese Academy of Sciences. \nc{We are grateful to the anonymous referee for careful and insightful comments that improved this work.} Xun Shi acknowledges helpful discussions with Kejia Lee and Yulan Liu, and thanks James McKee for discussions regarding observation targets. This work is supported by the National Natural Science Foundation of China grant No. 12373025. Yuanshang Huang is supported by Scientific Research and Innovation Project of Postgraduate Students in the Academic Degree of Yunnan University and the Project of Yunnan Provincial Department of Education Science Research Fund 
(Grant No. KC-252513932). Jumei Yao was supported by the National Science Foundation of Xinjiang Uygur Autonomous Region (2022D01D85), the Major Science and Technology Program of Xinjiang Uygur Autonomous Region (2022A03013-2), the Tianchi Talent Project, the CAS Project for Young Scientists in Basic Research (YSBR-063), the Tianshan Talents Program (2023TSYCTD0013), and the Chinese Academy of Sciences (CAS) ``Light of West China'' Program (No. xbzg-zdsys-202410 and No. 2022-XBQNXZ-015). Weiwei Zhu is supported by the National SKA Program of China (No. 2020SKA0120200), and the National Natural Science Foundation grant (No. 12041303). Yonghua Xu is supported by the National Natural Science Foundation (grant No. 12573104), the National Key R\&D Program of China (grant No. 2022YFC2205203), and the CAS ``Light of West China'' Program. 

\facility{FAST:500m}.

\software{AstroPy \citep{2013A&A...558A..33A, 2018AJ....156..123A, 2022ApJ...935..167A}, corner \citep{corner}, DSPSR \citep{van_Straten_2011}, emcee \citep{2013PASP..125..306F}, lmfit \citep{2016ascl.soft06014N}, SciPy \citep{2020SciPy-NMeth}, Matplotlib \citep{Hunter:2007} and NumPy \citep{harris2020array}}.


%




\bibliography{sample631}{}
\bibliographystyle{aasjournal}

\appendix

\section{Testing for arc apex offsets} \label{Testing for arc apex offsets}
\nc{
Two mechanisms can cause the apex of a parabolic arc to be offset from the origin: (1) the scintillation screen has an electron density gradient transverse to the line of sight, or (2) the scattered images do not lie along the line of sight. However, the arcs we analyze show no measurable apex offset from the origin, indicating that the screens in our study do not exhibit significant density gradients and that their morphologies are consistent with the ``parallel stripes model'' \citep{Shi_2021}.
}
\nc{
To illustrate that the scintillation arc apexes in our study show no significant offset from the origin, we consider the examples shown in Figures \ref{fig:B1237+25_20200118_9_arc}, \ref{fig:B1237+25_20230718_7_arc}, \ref{fig:B1842+14_20221020_all_arc}, and \ref{fig:B2021+51_20220917_all_arc} (corresponding to observations of PSR B1237+25 on 2020 January 17, PSR B1237+25 on 2023 July 18, PSR B1842+14 on 2022 October 20, and PSR B2021+51 on 2021 September 17, respectively). The curvature measurement uncertainties for the main arcs in these four figures (the G arc for PSR B1237+25 and the E arc for PSR B2021+51) are 3.0, 3.7, 92, and 78 $\mathrm{m}^{-1}\,\mathrm{mHz}^{-2}$, respectively (see Table \ref{tab:B1237+25_eta}, \ref{tab:B1842+14_eta}, and \ref{tab:B2021+51_eta}). We performed tests without reflection padding; instead, a Hanning window was applied to the outer 10\% of each edge before computing the FFT, while accounting for arc asymmetry in the fits. The differences between the most probable curvature values obtained from this alternative approach and those from reflection padding are 0.3, 0.5, 12, and 16 $\mathrm{m}^{-1}\,\mathrm{mHz}^{-2}$, respectively. These differences are smaller than the curvature measurement uncertainties.
}

\section{Estimating bow shock radii} \label{bow shock radii}
\nc{
We calculate the bow shock stand-off radius using the formula provided by \cite{2024MNRAS.527.7568O} (based on \citealt{Wilkin_1996} and \citealt{Chatterjee_2002}):
}
\begin{eqnarray}
\begin{aligned}
R_{0} & = \sqrt{\frac{\dot{E}}{4 \pi c \rho v_{*}^{2}}} \\
& \approx 225\,\mathrm{au} \times \left[\left(\frac{\dot{E}}{10^{33}\,\mathrm{erg\,s}^{-1}}\right)\right.\\
& \quad\quad\quad \left.\times\left(\frac{n_{H}}{\mathrm{cm}^{-3}}\right)^{-1}\left(\frac{v_{*}}{100\,\mathrm{km\,s}^{-1}}\right)^{-2}\right]^{1/2},
\label{R0}
\end{aligned}
\end{eqnarray}
\nc{
\noindent where $\dot{E}$ is the pulsar spin-down luminosity, $c$ is the speed of light, $\rho$ is the ISM density, $v_{*}$ is the pulsar velocity, and $n_{H}$ is the ISM hydrogen number density. $\dot{E}$ is given by $\dot{E} = 4\pi^{2}I\dot{P}/P^3$, where $I \equiv 10^{45}\,\mathrm{g\,cm}^2$ is the pulsar moment of inertia. The spin-down luminosities of PSR B1237+25 and PSR B2021+51 are $1.4 \times 10^{31}\,\mathrm{erg\,s}^{-1}$ and $8.2 \times 10^{32}\,\mathrm{erg\,s}^{-1}$, so they are unlikely to have observable bow shocks.
}
\nc{
Assuming bow shocks are present, we use the formula from \cite{Chatterjee_2002} describing the shape of a thin-shell bow shock to calculate the assumed bow shock radius along the line of sight:
}
\begin{eqnarray}
\begin{aligned}
R(\theta)=R_{0} \csc \theta \sqrt{3(1-\theta \cot \theta)}.
\label{R}
\end{aligned}
\end{eqnarray}
\nc{
\noindent This equation is in polar coordinates, where $R(\theta)$ is the bow shock radius, and $\theta$ is the polar angle from the axis along the pulsar velocity direction. We assume that the angle between the pulsar velocity direction and the sky plane $\alpha$ is between -50° and 50°, in which case the pulsar velocity is $v_{*} = v_{\perp} / \cos\alpha$ (where $v_{\perp}$ is the pulsar's transverse velocity), and $\theta$ ranges from 40° to 140°. We also assume that the ISM hydrogen number density $n_{H}$ ranges from $0.001$ to $0.1\,\mathrm{cm}^{-3}$. Using the above assumptions, we obtain from Equations \ref{R0} and \ref{R} that the bow shock radius along the line of sight would be between 19 AU and 912 AU for PSR B1237+25, and 571 AU to 27563 AU for PSR B2021+51. However, we can only know that the closest screen to PSR B1237+25 is at a distance no greater than $\sim$ 40 pc (see Table \ref{tab:parameters_screens}), while that for PSR B2021+51 is no greater than $\sim$ 70 pc (see Table \ref{tab:parameters_screens}), so we cannot determine whether the screens are at the bow shocks. Although our estimated range for the bow shock radius is very large due to the substantial uncertainties in the assumptions, it still tells us that the bow shock radius is much smaller than the uncertainty in the distance from the screen to the pulsar.
}
\nc{
If the screen is very close to the pulsar ($s \ll 1$), the curvature of the scintillation arc can be approximated from Equations \ref{vsr_iso} and \ref{eta_lambda} as:
}
\begin{equation}
\eta \approx  \frac{d_{\text{screen-psr}}}{2 ({v_{\mathrm{psr},\zeta} - v_{\mathrm{screen},\zeta}})^2},
\label{estimate_eta}
\end{equation}
\nc{
\noindent where $d_{\text{screen-psr}}$ is the distance between the screen and the pulsar, and $v_{\mathrm{psr},\zeta}$ and $v_{\mathrm{screen},\zeta}$ are the projections of the pulsar velocity and screen velocity onto the main scattering direction of the screen, respectively. If the closest screen to PSR B1237+25 is at the bow shock, we obtain from Equation \ref{estimate_eta} that $v_{\mathrm{psr},\zeta} - v_{\mathrm{screen},\zeta}$ is no greater than $5\,\mathrm{km\,s}^{-1}$, while for the closest screen to PSR B2021+51 it is no greater than $4\,\mathrm{km\,s}^{-1}$.
}

\section{Supplementary figures} \label{Supplementary figures}
\begin{figure}[ht]
\centering
\includegraphics[width=\columnwidth]{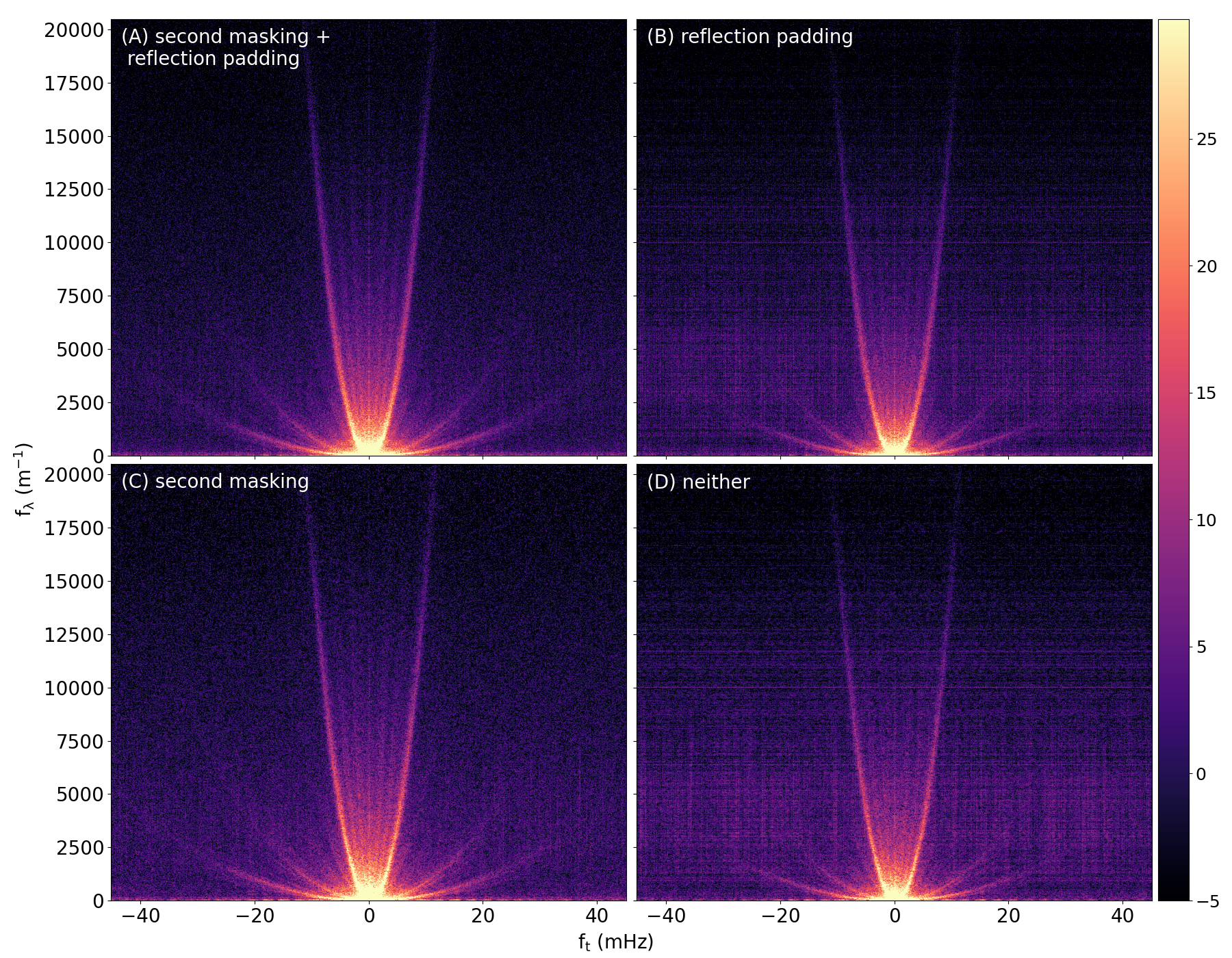}
\caption{Secondary spectra of PSR B1237+25 on January 17, 2020, showing the effects of different preprocessing methods applied to the dynamic spectrum. Panel A: secondary spectrum obtained from the dynamic spectrum with both second masking and reflection padding applied. Panel B: only reflection padding applied. Panel C: only second masking applied. Panel D: neither second masking nor reflection padding applied. For the dynamic spectra without reflection padding, a Hanning window was applied to the outer 10\% of each edge. We compare the background noise levels of each panel using the mean power away from the arc regions minus the 99.9th percentile power value: Panel A (second masking + reflection padding): -40.2 dB; Panel B (reflection padding only): -36.6 dB; Panel C (second masking only): -38.0 dB; Panel D (neither applied): -33.7 dB. Both second masking and reflection padding reduce the background noise in the resulting secondary spectrum.}
\label{fig:B1237+25_20200117_compare}
\end{figure}

\begin{figure}[ht]
\centering
\includegraphics[width=\columnwidth]{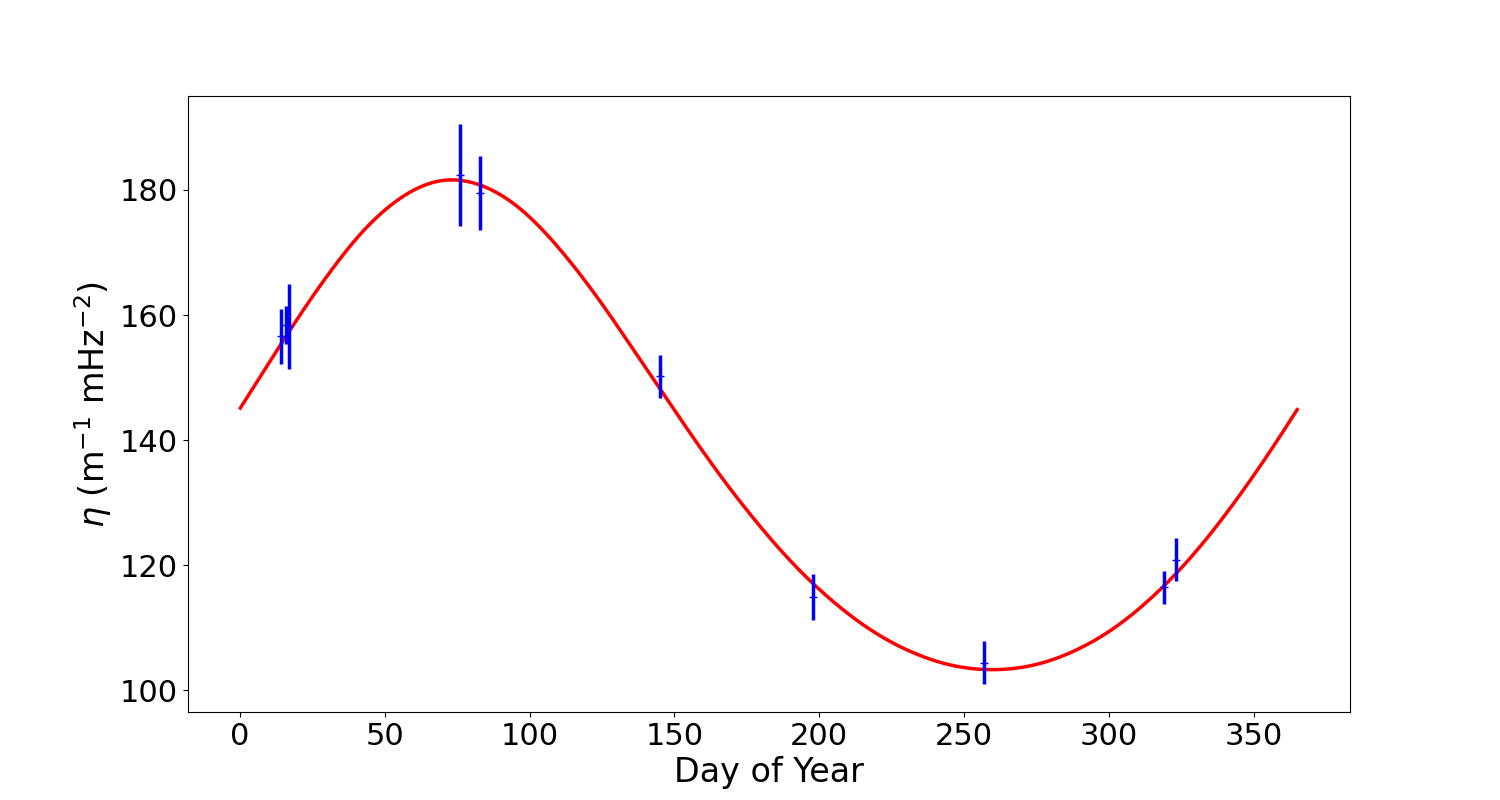}
\caption{Measurement results and errors of scintillation arc G curvature for PSR B1237+25 (blue lines) and the modeled annual modulation results (red curve). The vertical axis shows the scintillation arc curvature, while the horizontal axis represents the day of year corresponding to the observation time.}
\label{fig:B1237+25_G_eta_MCMC}
\end{figure}

\begin{figure}[ht]
\centering
\includegraphics[width=\columnwidth]{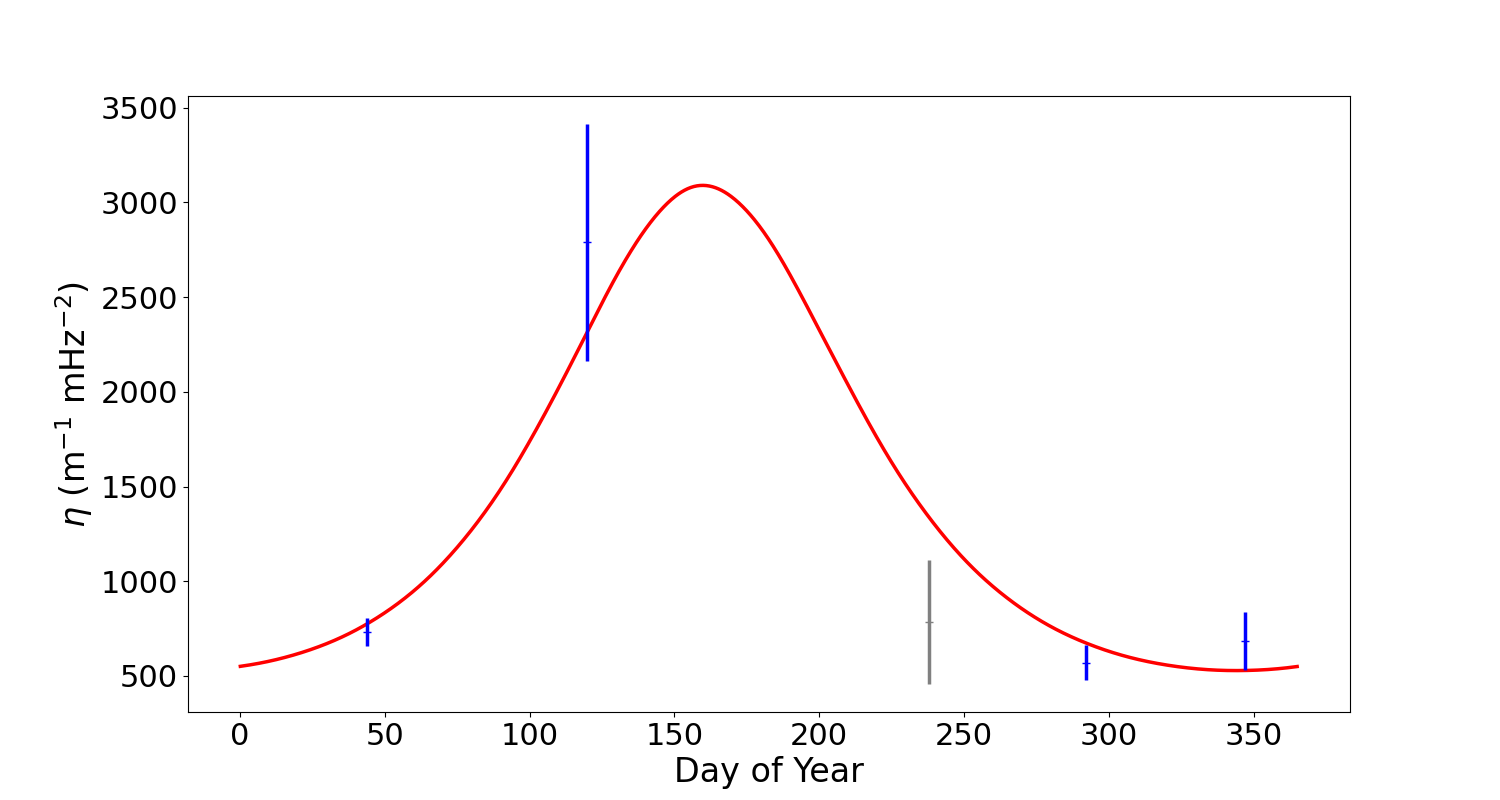}
\caption{Measurement results and errors of the scintillation arc curvature for PSR B1842+14 (blue lines) and the modeled annual modulation results (red curve). The measurement from \cite{2020RAA....20...76Y} is also shown (gray line). The vertical axis shows the scintillation arc curvature, while the horizontal axis represents the day of year corresponding to the observation time.}
\label{fig:B1842+14_eta_MCMC}
\end{figure}

\begin{figure}[ht]
\centering
\includegraphics[width=\columnwidth]{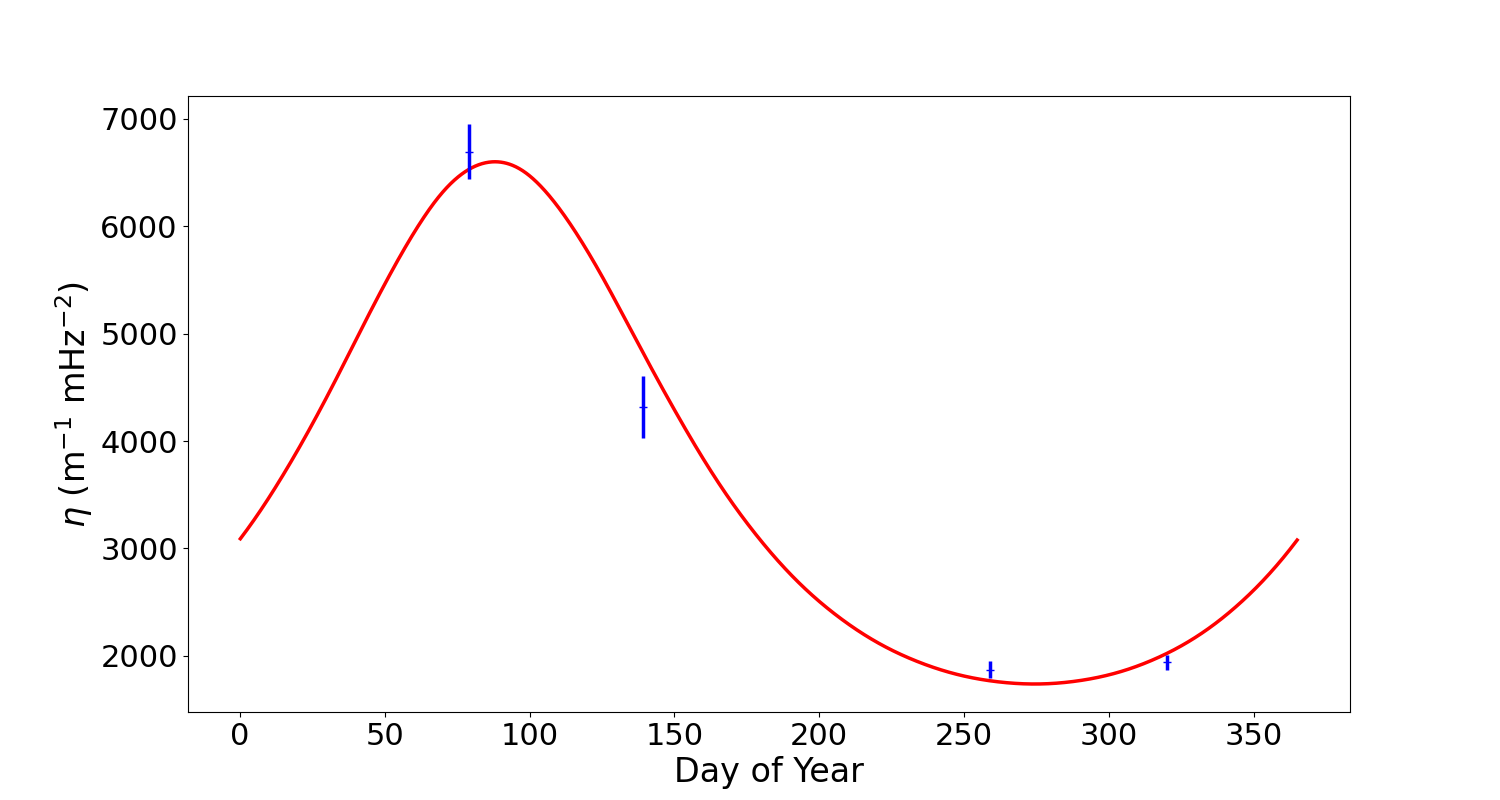}
\caption{Measurement results and errors of the scintillation arc E curvature for PSR B2021+51 (blue lines) and the modeled annual modulation results (red curve). The vertical axis shows the scintillation arc curvature, while the horizontal axis represents the day of year corresponding to the observation time.}
\label{fig:B2021+51_E_eta_MCMC}
\end{figure}

\begin{figure}[ht]
\centering
\includegraphics[width=\columnwidth]{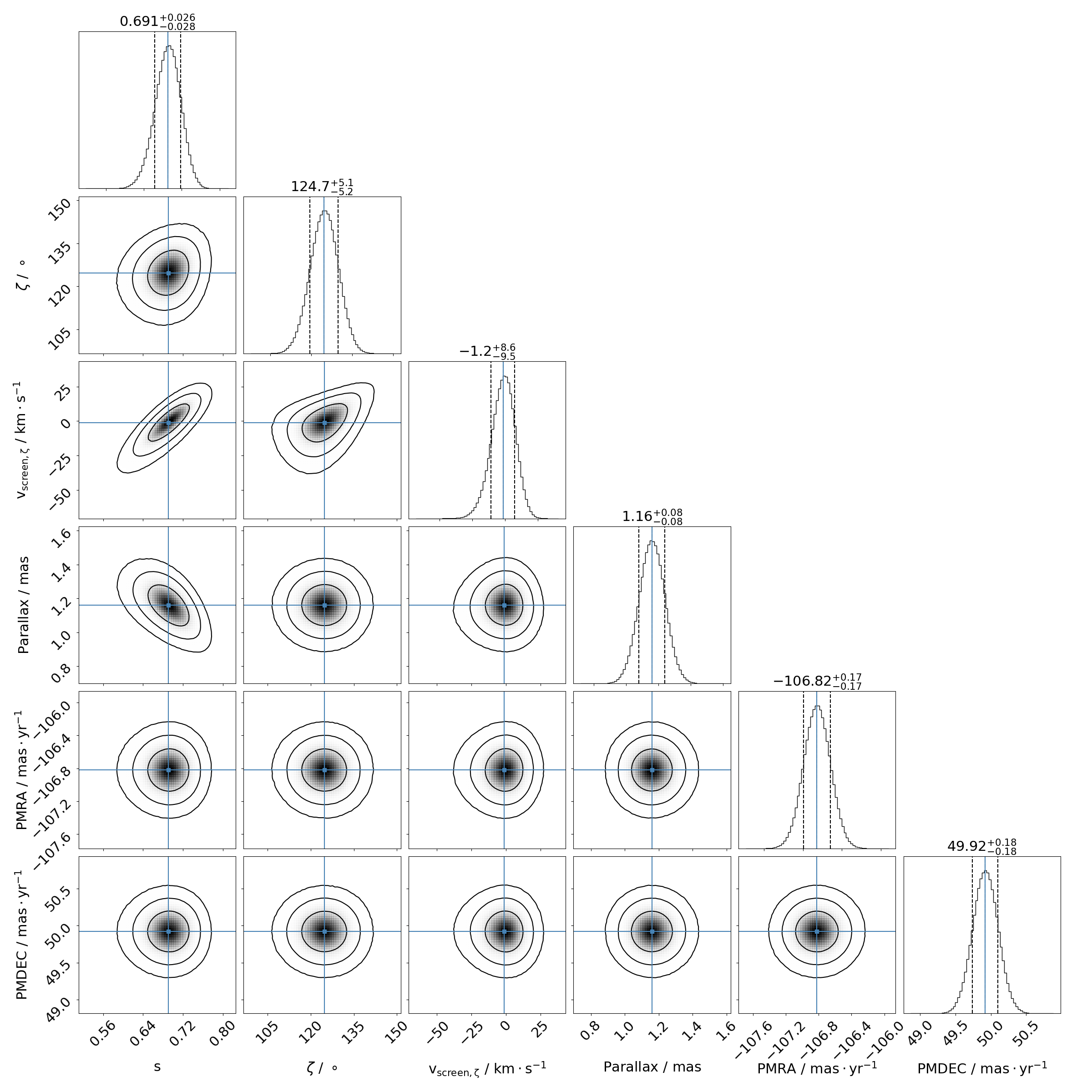}
\caption{The posterior distributions of parameters for screen G in the direction of PSR B1237+25, as well as the pulsar parallax and proper motion. In these figures, blue solid lines indicate the median, vertical dashed lines indicate 1 $\sigma$ confidence levels, and contours show 1 $\sigma$, 2 $\sigma$, and 3 $\sigma$ confidence levels.}
\label{fig:B1237+25_G_corner_full}
\end{figure}

\begin{figure}[ht]
\centering
\includegraphics[width=\columnwidth]{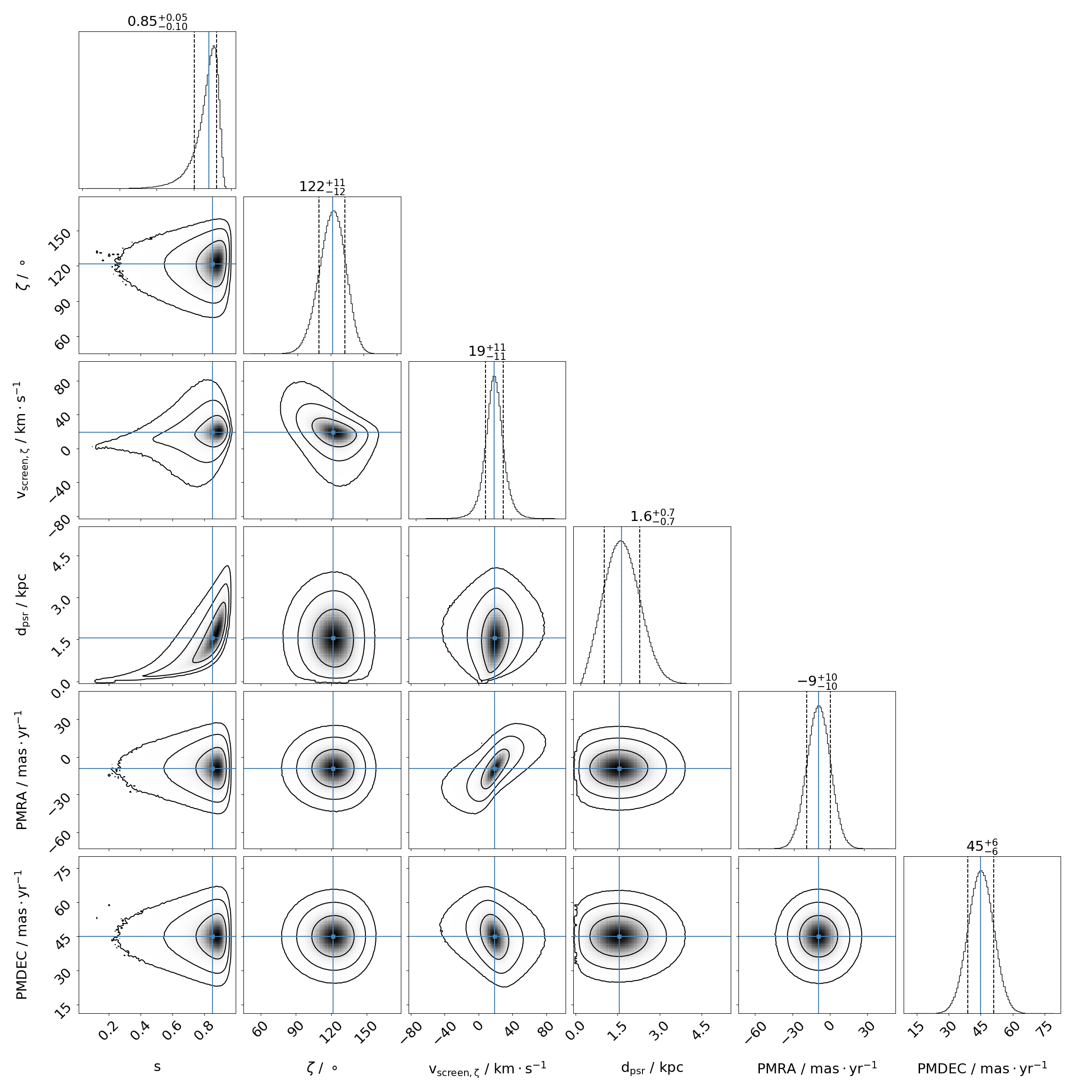}
\caption{The posterior distributions of parameters for the screen in the direction of PSR B1842+14, as well as the pulsar parallax and proper motion. In these figures, blue solid lines indicate the median, vertical dashed lines indicate 1 $\sigma$ confidence levels, and contours show 1 $\sigma$, 2 $\sigma$, and 3 $\sigma$ confidence levels.}
\label{fig:B1842+14_corner_full}
\end{figure}

\begin{figure}[ht]
\centering
\includegraphics[width=\columnwidth]{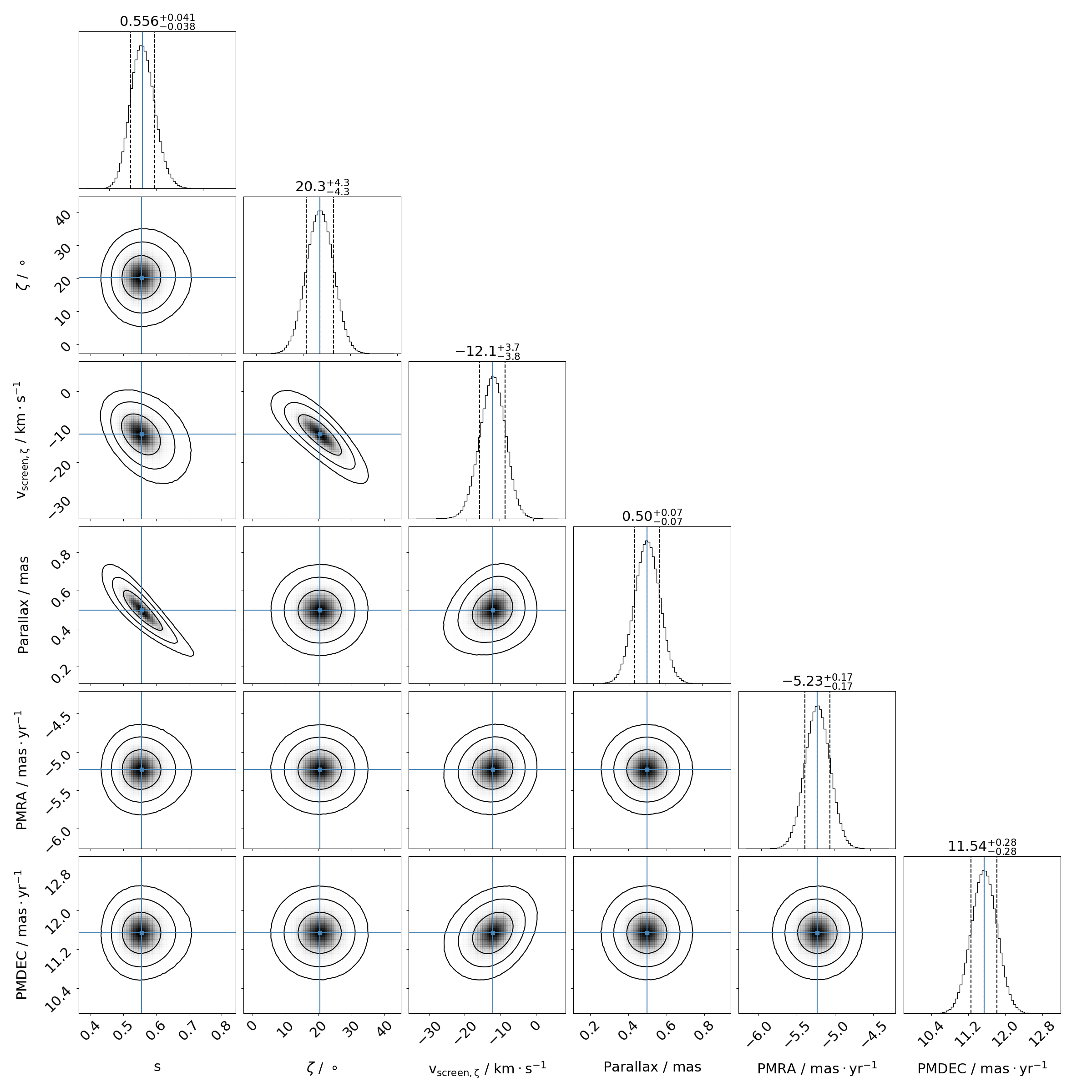}
\caption{The posterior distributions of parameters for screen E in the direction of PSR B2021+51, as well as the pulsar parallax and proper motion. In these figures, blue solid lines indicate the median, vertical dashed lines indicate 1 $\sigma$ confidence levels, and contours show 1 $\sigma$, 2 $\sigma$, and 3 $\sigma$ confidence levels.}
\label{fig:B2021+51_E_corner_full}
\end{figure}

\nc{
Figure \ref{fig:B1237+25_20200117_compare} illustrates the effectiveness of the second masking and reflection padding methods. 
Figures \ref{fig:B1237+25_G_eta_MCMC}, \ref{fig:B1842+14_eta_MCMC}, and \ref{fig:B2021+51_E_eta_MCMC} show the MCMC fitting results for the brightest arcs of pulsars B1237+25, B1842+41, and B2021+51, respectively. Figures \ref{fig:B1237+25_G_corner_full}, \ref{fig:B1842+14_corner_full}, and \ref{fig:B2021+51_E_corner_full} present the posterior distributions of the scintillation screen parameters that produce the brightest arcs of pulsars B1237+25, B1842+41, and B2021+51, respectively, as well as the posterior distributions of the pulsar parallax and proper motion.
}

\end{document}